\documentclass[%
 amsmath,amssymb,
 superscriptaddress,
]{revtex4-2}
\usepackage{soul,color}
\usepackage{multirow}
\usepackage{xcolor,colortbl}
\usepackage{hyperref}
\hypersetup{
    colorlinks=true,
    linkcolor=blue,
    filecolor=magenta,      
    urlcolor=cyan,
}

\usepackage{graphicx}% Include figure files
\usepackage{dcolumn}% Align table columns on decimal point
\usepackage{bm}% bold math
\begin{document}

\preprint{APS/123-QED}

\title{High-performance, additively-manufactured atomic spectroscopy apparatus for portable quantum technologies}

  \author{S H Madkhaly}
%\email{First.Author@institution.edu.}
 \affiliation{%
School of Physics and Astronomy, University of Nottingham, University Park, Nottingham, NG7 2RD, UK%\\This line break forced% with \\
}%
 \affiliation{%
 Department of Physics, Jazan University, Jazan, Kingdom of Saudi Arabia}%Lines break automatically or can be forced with \\
\author{N Cooper}%
\email{nathan.cooper@nottingham.ac.uk}
\affiliation{%
School of Physics and Astronomy, University of Nottingham, University Park, Nottingham, NG7 2RD, UK%\\This line break forced% with \\
}%

\author{L Coles}%
%\email{Second.Author@institution.edu.}
\affiliation{ Added Scientific Ltd, Unit 4, Isaac Newton Centre, Nottingham, NG7 2RH, UK}
%School of Physics and Astronomy, University of Nottingham, University Park, Nottingham, NG7 2RD, UK%\\This line break forced with \textbackslash\textbackslash
%}%

\author{L Hackerm\"{u}ller}
 \email{lucia.hackermuller@nottingham.ac.uk}
\affiliation{%
School of Physics and Astronomy, University of Nottingham, University Park, Nottingham, NG7 2RD, UK%\\This line break forced% with \\
}%

\begin{abstract}
\vskip  5mm
\hspace{6cm}{\textbf{Abstract}} \\
We demonstrate a miniaturised and highly robust system for performing Doppler-free spectroscopy on thermal atomic vapour for three frequencies as required for cold atom-based quantum technologies. The application of additive manufacturing techniques, together with efficient use of optical components, produce a compact, stable optical system, with a volume of 0.089\,L and a weight of 120\,g. The device occupies less than a tenth of the volume of, and is considerably lower cost than, conventional spectroscopic systems, but also offers excellent stability against environmental disturbances. We characterise the response of the system to changes in environmental temperature between 7 and 35\,$^\circ$C and exposure to vibrations between 0 - 2000\,Hz, finding that the system can reliably perform spectroscopic measurements despite substantial vibrational noise and temperature changes. Our results show that 3D-printed optical systems are an excellent solution for portable quantum technologies.
\end{abstract}

\maketitle

\section{Introduction}
Experiments involving cold atoms allow high-precision tests of fundamental physics \cite{YverF2003Acai, geiger2020high, tino2021testing, cacciapuoti2020testing, takamoto2020test, kaiser2019}, simulation of condensed matter systems \cite{HagueJP2014CRAf, lewenstein2007ultracold, demler2014strongly} and the study of new states of matter \cite{DalfovoFranco1999ToBc, LeggettAnthonyJ2001Bcit, BlochImmanuel2008Mpwu, GiorginiStefano2008Toua}. They also offer one of the most promising routes towards quantum information processing \cite{farrera2016generation, Garcia-RipollJJ2005Qipw, RyabtsevII2016Socr} and quantum communication \cite{roztocki2017, schneeweiss2018cold, kuzmich2003generation, bao2012efficient}. Furthermore, the techniques used in these experiments underpin an important emerging field of technological development, in the form of atomic quantum sensors and cold atom clocks \cite{bongs2014isense, falke2014strontium, ludlow2015optical, riehle2017optical}. 

These experiments require one or more lasers to be frequency-stabilised via feedback based on thermal vapour spectroscopy, and most require the production of a magneto-optical trap (MOT) \cite{zeeman1}. The frequency stability of these lasers is key to the performance of the experiment, yet the optical systems used for vapour spectroscopy are often not well-optimised, bulky and occupy a large fraction of an optical table. Recently, the use of additive manufacturing (AM, also known as 3D printing) methods, together with novel approaches to experimental design, has been demonstrated to enable a compact magneto-optical trap system that captures $>\,2\times 10^8$ atoms \cite{Cooper2021, optamot1}. 
AM techniques allow devices to be fully optimised for their intended function, without reference to traditional design constraints \cite{jared2017additive}.
\begin{figure}[htbp!]
\centering
    \includegraphics[width=0.8\textwidth]{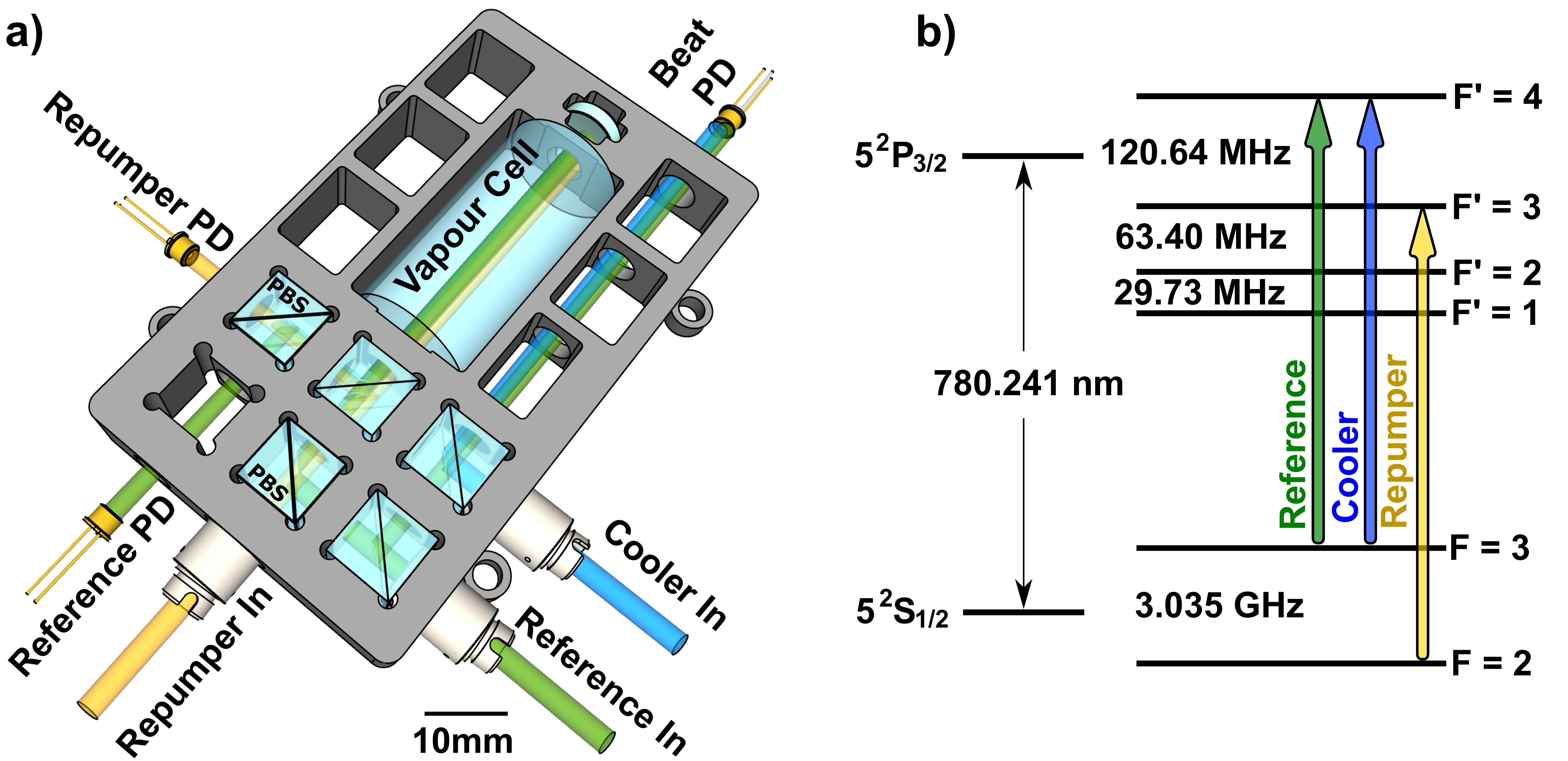}
    \caption{a) Compact system for performing spectroscopic measurements on three lasers simultaneously, exploiting a monolithic, additively manufactured frame. b) Transition structure of the $^{85}$Rb D2 line, indicating the names traditionally given to lasers used for magneto-optical trapping and cold atom experiments, as well as the transitions they address.}
    \label{allamo}
\end{figure}

Herein, we apply these broader principles
to the construction of a separate, specialised device for laser frequency stabilisation and characterise its performance under environmental stress such as temperature changes and vibrations.
While the specific prototype device we demonstrate as an example is aimed at applications within atomic physics, the techniques and design principles employed are transferable and may be of use in any field relying on hitherto complex free-space optics.

Our methods have been deliberately chosen so as to facilitate replication and augmentation of this approach: we focus on low-cost, widely-available hardware components and open source software, while keeping the complexity of the techniques and devices employed to a minimum. Our AM-based approach enables remarkable stability and robustness against environmental changes and makes devices constructed along these lines a cost-effective option for deployment in quantum technologies. For portable quantum technologies \cite{mcgilligan2017grating, rocket2017jokarus, strangfeld2021prototype} and fundamental experiments in space \cite{belenchia2022quantum, frye2021bose, abou2020aedge, elliott2018nasa}, stability with respect to temperature changes and vibrations is of key importance. The device demonstrated here also employs an unconventional spectroscopy scheme, in which light resonant with two different atomic transitions is spatially-overlapped in the atomic vapour, that has been found to offer some advantages in terms of signal strength and sensitivity \cite{cooper2021dual}.  
 The different features and components of the device are discussed in more detail below. 

\section{Monolithic optomechanical framework}
\label{AllamoI}
Our prototype device consists of a monolithic, AM framework that is populated with off-the-shelf optical components, as seen in Fig.\,\ref{allamo}.

\begin{figure}[ht]
\centering
    \includegraphics[width=0.5\textwidth]{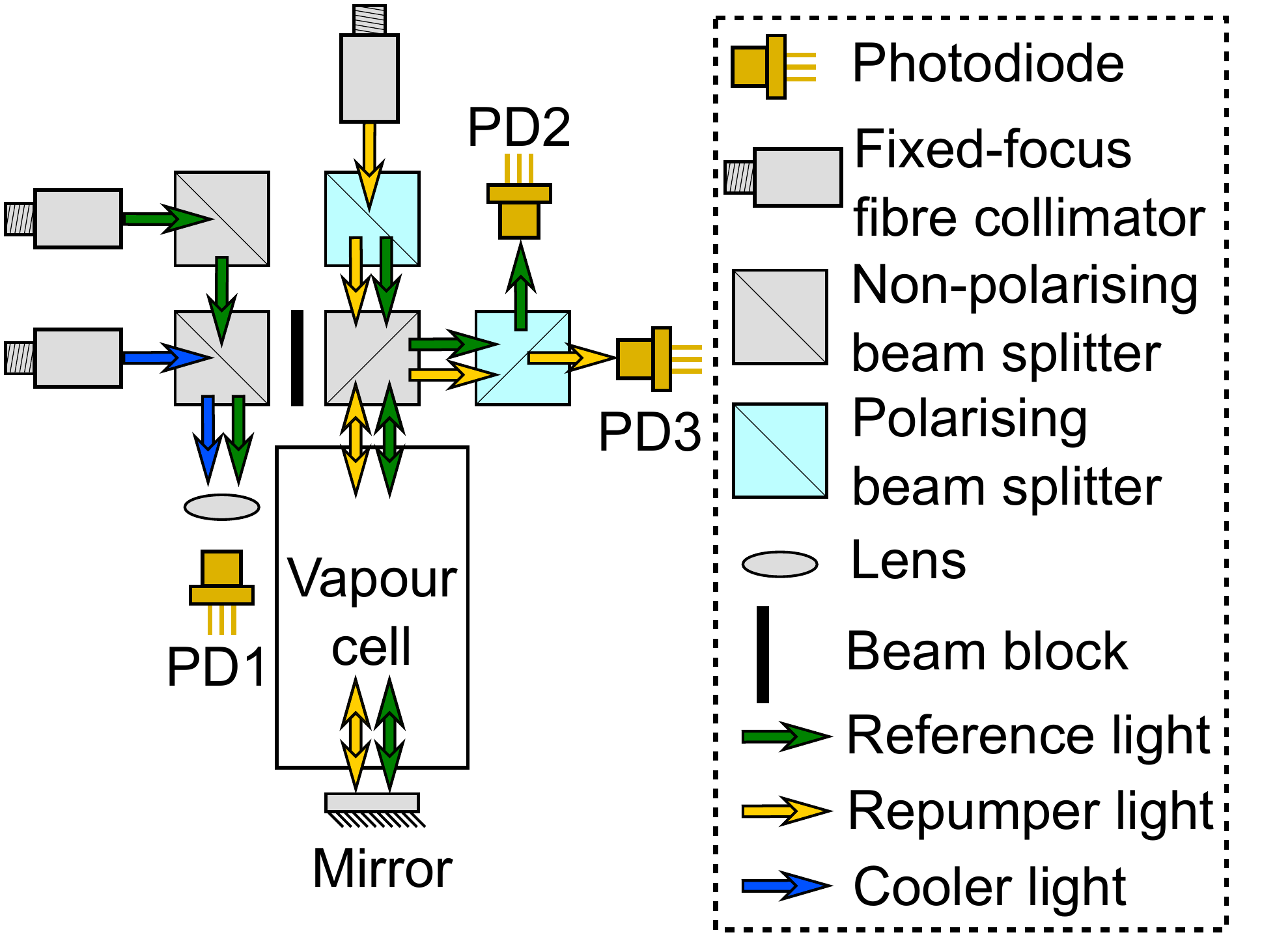}
    \caption{Schematic diagram showing the optical pathways within the compact spectroscopy device.}
    \label{optdiag}
\end{figure}

This framework was additively manufactured from polylactic acid via Fused Deposition Modelling (FDM) printing \cite{ultimaker, simplify3d, friedrich2020structure} and designed to house a minimal optical setup for
simultaneous frequency stabilisation of three lasers via saturated absorption spectroscopy \cite{sas}. The total print time was about 3 hours with no post processing required. A model of the framework, with optical components implanted, is shown in Fig.\,\ref{allamo}, and the relevant beam pathways are shown schematically in Fig.\,\ref{optdiag}.

The device has outer dimensions of 110$\times$65$\times$12.5\,mm and a total weight of 120\,g (including optics). 
The design of the framework keeps optical path lengths as short as possible to improve stability. The maximum optical path length for any of the beams in the device is 230\,mm, while a typical path length in a standard experimental apparatus would be on the order of 1-2\,m. Stability is also achieved by omitting adjustable elements. 

The optical components push-fit directly into the polymer framework, exploiting AM design features described in more detail in \cite{optamot1}; holes in the framework allow the passage of the laser beams where required and recesses over the central regions of the optics allow them to be slotted into the framework without risk of scuffing the optically-active surfaces. Small holes in the corners of the component slots improve push-fit alignment by favouring an extended contact region rather than a point contact, and also avoid the build defects that can otherwise result from filament dragging \cite{goh2020process}. A grid-like structure adds stability, reduces material and supports a homogeneous printing process. The layout of the beam paths is shown in Fig.\,\ref{optdiag}.

\section{Optical components and pathways, control electronics}
The layout of the spectroscopy apparatus is shown in Fig.\,\ref{allamo}\,(a). 
All components are of industry standard sizes and the device can be reproduced with components from a range of manufacturers. Fig.\,\ref{optdiag} shows the schematic outline of the optical pathways within the device, indicating how light from each of the three input connectors is employed to generate the required spectroscopic signals and optical beat note. The photodiodes are operated using homebuilt transimpedance amplifiers with buffered outputs. The LM324N quadruple operational amplifier chip is employed for the spectroscopic photodiodes and the (higher bandwidth) AD8001A operational amplifier is used for the optical beat note (note that some data was collected using commercial photodiode-amplifier packages, prior to the construction of the homebuilt models --- see e.g. Fig.\,\ref{mechsetup}).

\begin{figure}[hb]
\centering
    \includegraphics[width=0.6\textwidth]{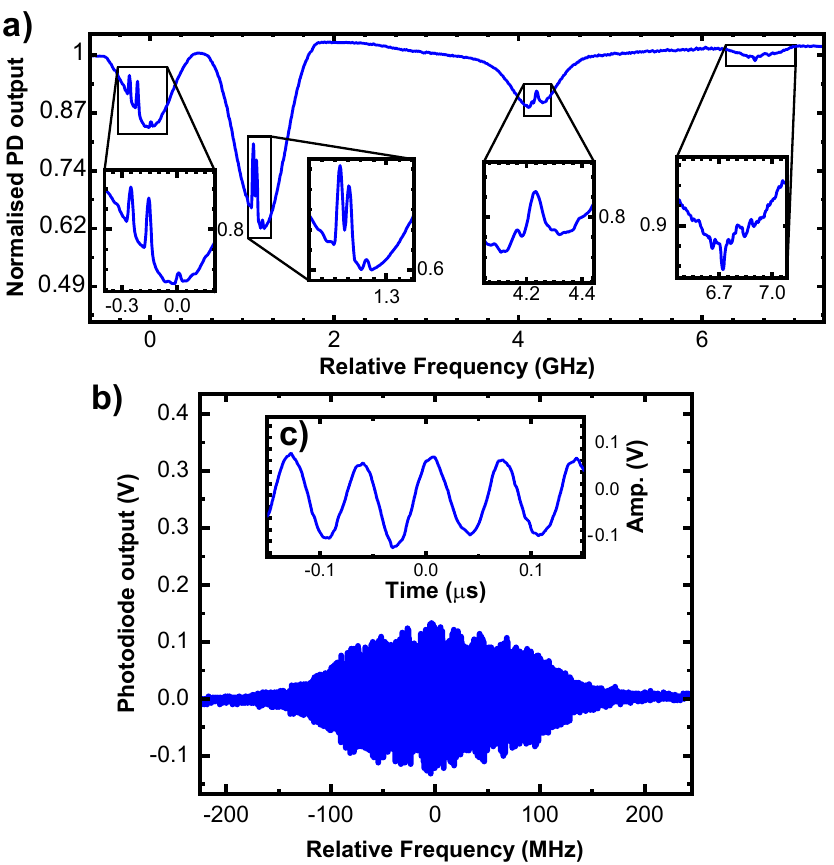} %13cm
    \caption{Saturated absorption spectra of $^{85}$Rb and $^{87}$Rb isotopes (a), and beat signal between a laser tuned to the $^{85}$Rb cooler transition and the reference laser (b), both obtained using the optical system illustrated in Fig.\,\ref{allamo}. The inset (c) displays a single-frequency beat note at ($13.4\pm$0.1)\,MHz.}
    \label{SASallamo}
\end{figure}

All beams are brought into the optical framework via fixed-focus fibre collimators and then distributed and routed via beam splitters. Fig.\,\ref{optdiag} shows a diagram of the beam paths. The reference beam is immediately split into two pathways: one for vapour spectroscopy and one to produce an optical beat note. The component used for the optical beat note is combined with the cooler light, at a non-polarising beam splitter, following which the two beams are overlapped on the photodiode `PD1.' The component to be used for spectroscopy is combined at a polarising beam splitter with the repumper light. The two beams then propagate through the vapour cell and back, with a portion of the return light picked-off using a non-polarising beam splitter; the reference and repumper frequency components of the return light are then separated via a polarising beam splitter and directed onto photodiodes `PD2' and `PD3' respectively. The double-pass beam configuration within the vapour cell \cite{cooper2021dual} allows for Doppler-free spectroscopy to be performed, resulting in the saturated absorption signal of D2 line of $^{85}$Rb and $^{87}$Rb shown in Fig.\,\ref{SASallamo} (a). The device also produces an optical beat note (Fig.\,\ref{SASallamo} (b)), on photodiode PD1, which allows the cooler laser to be stabilised with a controllable frequency separation from the reference laser.\\
Fig.\,\ref{SASallamo} (b) shows the beat note created between the reference laser and the cooler laser, obtained from the reference laser being stabilised to the $^{85}$Rb $F=3 \rightarrow F'=4$ transition and the cooler laser being scanned over the full range. The main panel shows the beat signal envelope as the cooler laser is scanned; this gives an indication of the frequency range and signal to noise ratio of the beat signal, but necessarily undersamples the individual beat frequencies present at any given point within the scan; the inset therefore displays an example of a single-frequency beat note at (13.4$\pm$0.1)\,MHz.% The frequencies contributing to Fig.\,\ref{SASallamo} (b) range from xxx to xxx, demonstrating a large range of available offset frequencies (200MHz) and a good signal to noise ration (~ factor 100?). Note, as a consequence of the large frequency range the data in Fig.\,\ref{SASallamo} (b) is slightly undersampled. The inset ...} 

To improve the laser frequency stability while reducing size, weight and cost, control electronics based on digital microcontrollers were utilised. Specifically, the Arduino Uno development board was used, in combination with its associated open-source software, as the central component of our control and feedback electronics. 

\begin{figure}[htbp]
 \centering
   \includegraphics[width=0.8\textwidth]{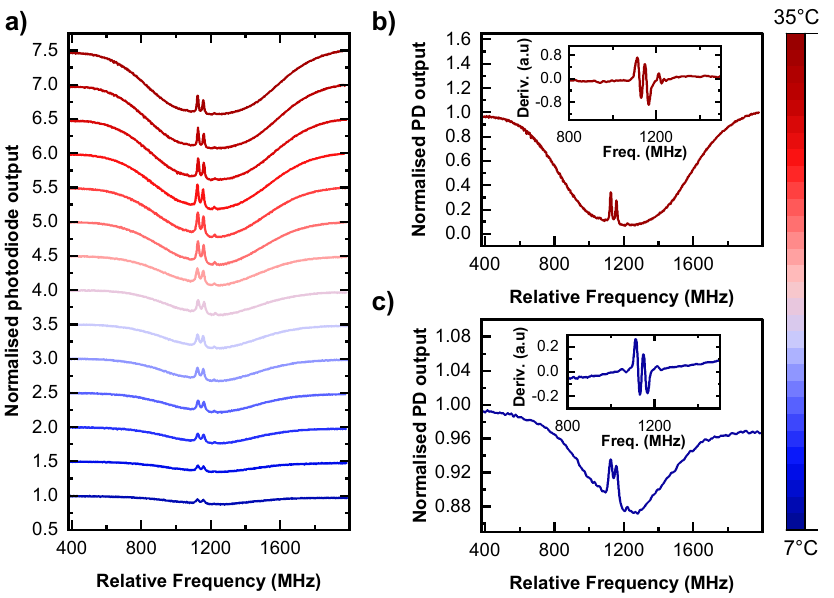} %13cm
   \caption{(a) Spectroscopic signal line shape of $^{85}$Rb $F=3 \rightarrow F'=2,3,4$ transition corresponding to temperature changes from 7$^{\circ}$\,C to 35$^{\circ}$\,C. For visibility, an offset of 0.5 is applied between successive traces. As the temperature rises from 7$^{\circ}$\,C (panel (c)) to 35$^{\circ}$\,C (panel (b)), the spectroscopic signal amplitude increases while its form remains unchanged. The insets in (b) and (c) represent the derivative signals of the corresponding traces.}
   \label{thermal}
   \end{figure}

The system has been tested using both external cavity diode lasers (e.g. for the data in figure \ref{mech1}) and distributed feedback lasers (e.g. as used for figures \ref{SASallamo}(a) and \ref{thermal}). A 1/e$^2$ beam diameter of 2.1\,mm was used and spectroscopy signals were obtained with beam powers ranging from 0.5 to 5\,mW. By dividing the r.m.s. voltage noise on the gradient of the spectroscopic signal (which is typically used as a feedback signal for laser frequency stabilisation) by its sensitivity to frequency variations about the laser lock point, we estimate the spectroscopic limit on laser linewidth using our current system, with a feedback bandwidth of 10\,kHz, to be (1.28$\pm$0.02)\,MHz. This would be more than sufficient for most applications, such as the creation of a magneto-optical trap (shown in \cite{optamot1}), though certain specialist applications may require narrower linewidths. These could be obtained at relatively modest cost through the use of superior photodiode amplifiers; alternatively, if the majority of the contributions to the free-running laser's frequency variations come from the low-frequency end of the spectrum, as is usually the case, then the spectroscopic output can be time-averaged or put through a low-pass filter to improve signal-to-noise by sacrificing bandwidth.  
 
\section{Stability measurements}
To ensure that our system is suitable for use outside the laboratory, a series of stability tests under harsh environmental conditions was performed.

 %\begin{figure}[htbp]
 %\centering
 %  \includegraphics[width=0.9\textwidth]{Fig4_error_signals.pdf} %13cm
 %  \caption{Spectroscopic signal line shape of $^{85}$Rb $F=3 %\rightarrow F'=2,3,4$ transition corresponding to temperature changes %from 7$^{\circ}$\,C to 35$^{\circ}$\,C (a). As the temperature rises %from 7$^{\circ}$\,C (c) to 35$^{\circ}$\,C (b), the spectroscopic %signal amplitude increases while its form remains unchanged. For %visibility, an offset of 0.5 is applied between successive traces.}
 %  \label{thermal}
 %  \end{figure} 

% \begin{figure}[htbp]
%  \centering
%   \includegraphics[width=0.9\textwidth]{Fig4_bc_insets.pdf} %13cm
%   \caption{Spectroscopic signal line shape of $^{85}$Rb $F=3 \rightarrow F'=2,3,4$ transition corresponding to temperature changes from 7$^{\circ}$\,C to 35$^{\circ}$\,C (a). As the temperature rises from 7$^{\circ}$\,C (c) to 35$^{\circ}$\,C (b), the spectroscopic signal amplitude increases while its form remains unchanged. \textcolor{red}{The insets in (b) and (c) represent the derivative signal of the corresponding trace.} For visibility, an offset of 0.5 is applied between successive traces.}
%   \label{thermal}
%   \end{figure} 

\subsection{Thermal Stability Test}
In order to test the optical system's thermal stability, it was subjected to changes in the environmental temperature while monitoring the resulting spectroscopic signals. The optical framework was placed inside an aluminium box to create a uniform thermal distribution around the system, and the temperature was gradually raised using a heating tape. This process was performed sufficiently slowly to ensure that the system reached thermal equilibrium prior to data collection. The change in temperature was monitored using three temperature and humidity sensors placed at different points inside and outside the metal box. The laser frequency was continuously scanned across the $F=3 \rightarrow F'=2,3, 4$  transitions of the $^{85}$Rb D2 line, in order to measure the system's response to the temperature changes. The spectroscopic signal was captured at a range of temperatures and is plotted in Fig.\,\ref{thermal}. 

As seen in the figure, the device performs its intended function over the full temperature range, from 7$^{\circ}$\,C to 35$^{\circ}$\,C. The form of the spectroscopic response is largely unchanged, while its amplitude increases at higher temperatures due to the increased vapour pressure of the alkali metal within the vapour cell. The spectroscopy traces for the highest (35$^{\circ}$\,C) and lowest temperatures (7$^{\circ}$\,C) are shown separately in Fig.\,\ref{thermal}\,(b), (c). The derivatives for these traces are given in the insets and depict an error signal with good signal-to-noise. The observed performance is at least comparable to that reported for considerably more expensive systems, made out of materials specifically selected for thermal stability, such as Invar and Zerodur \cite{duncker2014ultrastable}.

\begin{figure}[htbp]
 \centering
   \includegraphics[width=0.8\textwidth]{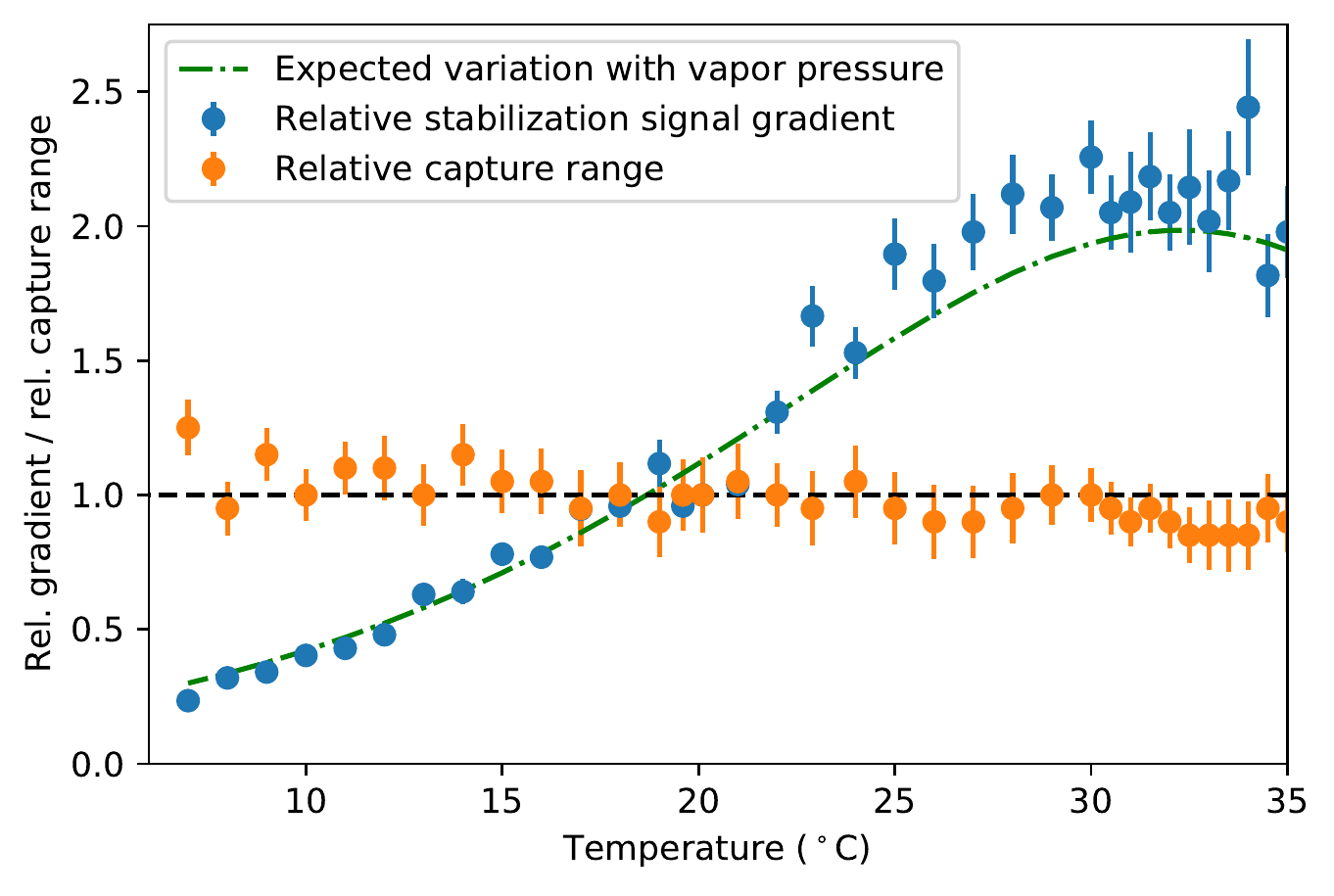} %13cm
   \caption{Relative sensitivity of laser stabilisation feedback signal (given by the maximum second derivative of the spectroscopic signal with respect to laser frequency) as a function of device temperature (blue data). The green line indicates the expected variation due to the change in Rb vapor pressure. %, and is fitted using one free parameter (a universal scaling coefficient by which all the values are multiplied). 
   The orange data shows the capture range. In both cases, normalization is relative to the values measured at 20\,$^\circ$C.}
   \label{errsig}
   \end{figure}

The derivatives according to frequency have been calculated for all traces shown in Fig.\,\ref{thermal}\,(a), together with additional data of the same nature that was omitted from Fig.\,\ref{thermal}\,(a) to avoid overcrowding, and were used to obtain a quantitative, temperature dependent comparison of the relative capture range and the relative stabilisation signal gradient.  Figure \ref{errsig} shows the variation of the sensitivity of a laser stabilisation feedback signal (corresponding to the gradient of the spectroscopic signal) to laser frequency (blue data points), normalized to the result at lab temperature (20$^{\circ}$\,C). In order to obtain these results, the first and second derivatives of the spectroscopic signals were obtained via a 3-point central difference method. Since the dominant source of error in the plotted data comes from high-frequency electrical noise in the photodiode amplifier circuit, such that the noise contributions even to adjacent sample points show very little correlation, multiple gradient estimates with different sample widths can be used to obtain an average value and associated statistical error estimate, which is plotted in the figure.      

The dash-dotted green line indicates the expected variation due to vapor pressure changes; we obtain this result by assuming that all points in the observed signal individually obey the Beer-Lambert law, and that the absorption coefficient for each laser frequency scales in proportion to the vapor pressure in the cell. This model is sufficient to yield most of the variation observed over the relevant parameter range.
It can be seen that our experimental data shows a slightly stronger temperature response, particularly in the higher temperature range. This can be attributed to the fact that our model neglects saturation effects; at high vapor pressure, the increased attenuation of the light reduces the saturation level, and consequent power broadening, that occurs at positions further along the optical path. This reduction in power broadening sharpens the sub-Doppler features slightly and hence can be expected to lead to a small increase in the gradient of the stabilisation signal, as observed. Importantly, our results do not indicate any degradation of the performance of the spectroscopic system over the full temperature window. 

Also plotted is the relative capture range of the stabilisation signal (orange data points); this is given by the frequency range about the desired stabilisation frequency within which the gradient of the feedback signal does not change sign, normalised to the result at lab temperature (20$^{\circ}$\,C). This parameter is a measure of the robustness of a laser stabilisation technique. Derivative and error determination relies on a 3-point central difference approach, together with standard error propagation to convert uncertainty in the values of the second derivative of the spectroscopic signal to uncertainty in the laser frequency at which it changes sign. The results show that the performance of the spectroscopic system is not degraded as the temperature moves away from the standard operating temperature of 20$^{\circ}$\,C. The slight downward trend in capture range as temperature is increased can be attributed to the same mechanism of reduced power-broadening as described above.  

\subsection{Response to Vibrations}
\begin{figure}[htbp]
\centering
\includegraphics[width=0.6\textwidth]{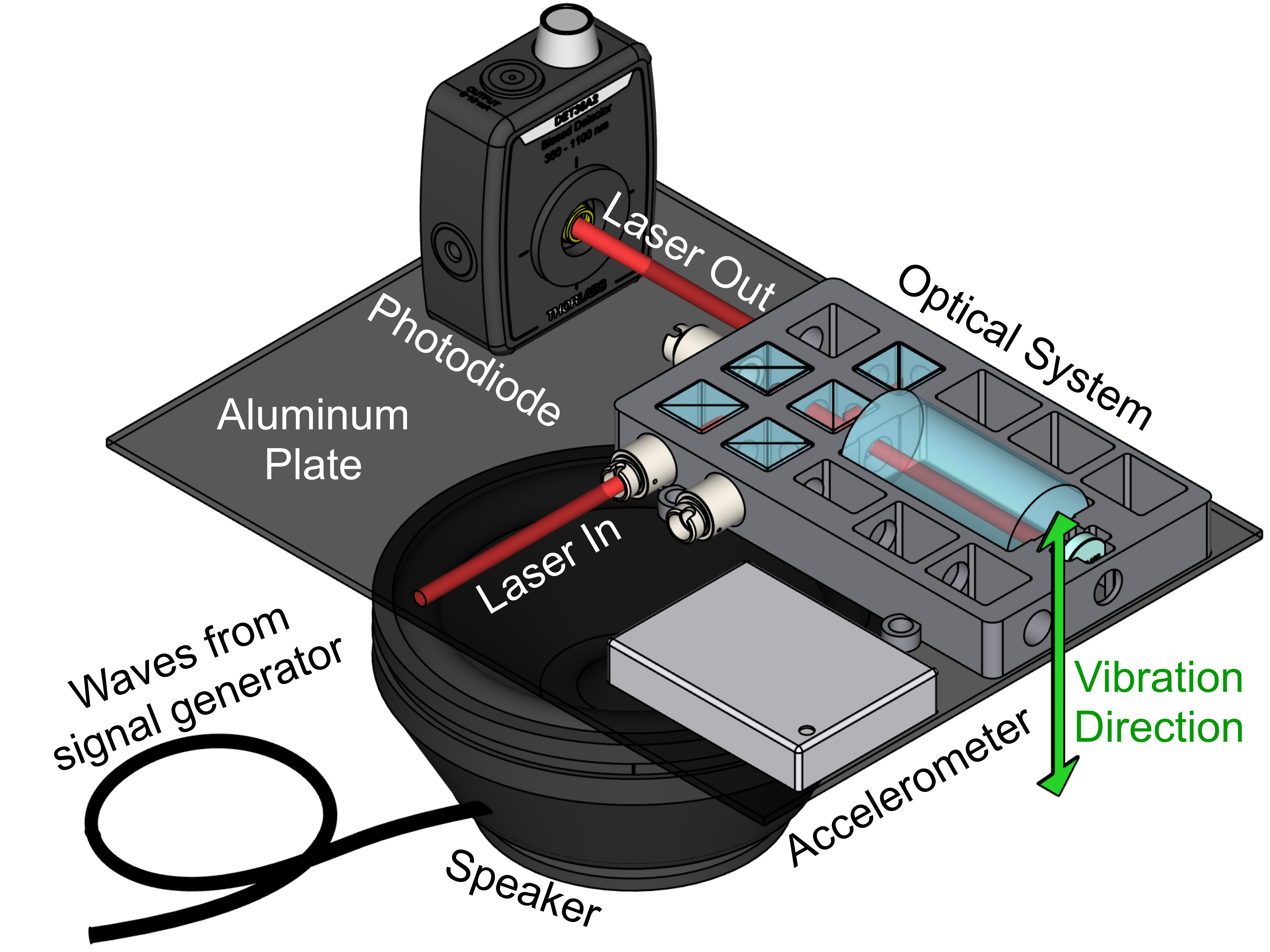} %13cm
 \caption{Mechanical stability test setup. A loudspeaker was used to generate a vibrating environment with varying frequencies to evaluate the optical framework's stability ---see text for details.}
\label{mechsetup}
\end{figure}

Mechanical instability is a major challenge that needs to be overcome when designing portable systems for space applications and work outside laboratory walls \cite{ressel2010ultrastable}. For this purpose, we simulated a vibrating environment by placing the device onto an aluminium plate, as shown in Fig.\,\ref{mechsetup}. 

An analog monitoring accelerometer (Te-connectivity 820M1), placed as shown in the figure, was used to measure and record the vibrations to which the device was exposed. Sinusoidal vibrations were introduced at several frequencies, between 0 Hz and 2 kHz, generated by a frequency generator and a loudspeaker. Additionally, manual shaking of the plate was used to simulate a noisy environment. 

%\begin{figure}[htbp]
%\centering
%\includegraphics[width=\textwidth]{CSA_Fig6_9.pdf} %13cm
% \caption{Measurements of the optical system's response to single-frequency vibrations (100Hz, 500Hz, 1kHz, 2kHz) generated from a loudspeaker (a,b, c, and d) and to manual shaking (e). The power spectral density (PSD) of the output signal from the reference photodiode (PD2) is shown as a function of frequency, both when the laser is  stabilised on resonance with the $^{85}$Rb $F=3 \rightarrow F'=4$ transition (green), and when the laser is free-running about the same frequency (blue). The PSD of the vibrations measured by 820M1-TE accelerometer (see Fig.\,\ref{mechsetup}) is also shown (orange), along with the PSD of the driving voltage applied to the loudspeaker (black).}
%\label{mech1}
%\end{figure}
\begin{figure}[htbp]
\centering
\includegraphics[width=\textwidth]{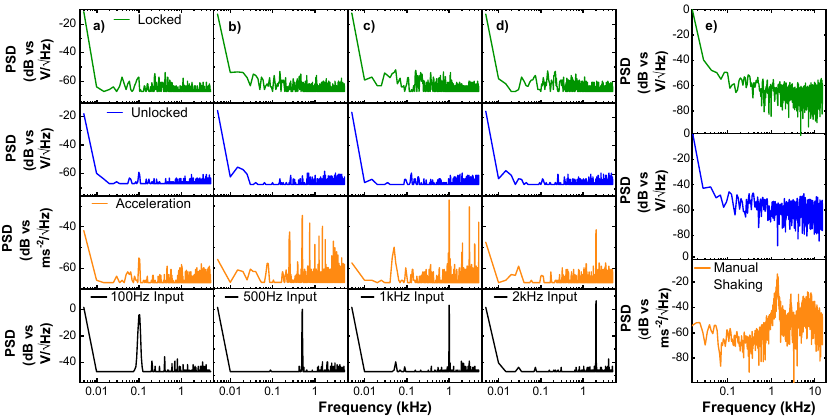} %13cm
 \caption{Measurements of the optical system's response to single-frequency vibrations (100Hz, 500Hz, 1kHz, 2kHz) generated from a loudspeaker (a,b, c, and d) and to manual shaking (e). The power spectral density (PSD) of the output signal from the reference photodiode (PD2) is shown as a function of frequency, both when the laser is  stabilised on resonance with the $^{85}$Rb $F=3 \rightarrow F'=4$ transition (green), and when the laser is free-running about the same frequency (blue). The PSD of the vibrations measured by 820M1-TE accelerometer (see Fig.\,\ref{mechsetup}) is also shown (orange), along with the PSD of the driving voltage applied to the loudspeaker (black).}
\label{mech1}
\end{figure}
%\begin{figure}[htbp]
%\centering
%\includegraphics[width=\textwidth]{CSA_Fig6_7.pdf} %13cm
% \caption{Measurements of the optical system's response to single-frequency vibrations (100Hz, 500Hz, 1kHz, 2kHz) generated from a loudspeaker (a,b, c, and d) and to manual shaking (e). The power spectral density (PSD) of the output signal from the reference photodiode (PD2) is shown as a function of frequency, both when the laser is  stabilised on resonance with the $^{85}$Rb $F=3 \rightarrow F'=4$ transition (green), and when the laser is free-running about the same frequency (blue). The PSD of the vibrations measured by 820M1-TE accelerometer (see Fig.\,\ref{mechsetup}) is also shown (orange), along with the PSD of the driving voltage applied to the loudspeaker (black).}
%\label{mech1}
%\end{figure}

 Figure \ref{mech1} displays the power spectral density of the control voltage used to drive the loudspeaker (black, lowest panel), the measured acceleration of the vibrating plate (orange), and the resulting photodiode output voltages from the reference photodiode, PD2, with the laser's frequency stabilisation active (green) and inactive (blue). For reference, the DC voltage level of the photodiode output voltages was $\sim$\,50\,mV. This is shown for driving frequencies of 100 Hz, 500 Hz, 1 kHz and 2 kHz (panels a) to d) respectively) as well as for manual shaking (panel e)). For sinusoidal, single-frequency vibrations, an amplitude of $\pm\,0.2$\,ms$^{-2}$ was applied, corresponding to the maximum output of the speaker. For the random shaking test (panel e)), vibrations were applied both manually and through the loudspeaker, with amplitudes going far beyond what was possible with the speaker alone; corresponding data was collected at frequencies up to 15\,kHz. 
 Despite clear contributions to the measured acceleration spectrum, very little contribution is made to the power spectrum of the optical signal by the presence of vibrations. 

 A requirement for a device to be suitable for portable quantum technologies is that it can withstand the acceleration and vibrations caused by transportation, e.g. those experienced in a moving vehicle \cite{chonhenchob2012measurement} or during a space launch \cite{schkolnik2016compact}, then operate in a stationary but un-isolated environment. Our data shows that the device described herein can operate well not only after, but also during, manual shaking, which produced acceleration amplitudes comparable to those experienced during high-speed transportation \cite{chonhenchob2012measurement,schkolnik2016compact}. Perturbation in an uncontrolled but stable environment outside the lab (e.g. standing on the ground or on a platform) are likely to be of similar amplitudes to what we have tested in Fig.\ref{mech1}

\begin{figure}[htbp]
\centering
\includegraphics[width=0.7\textwidth]{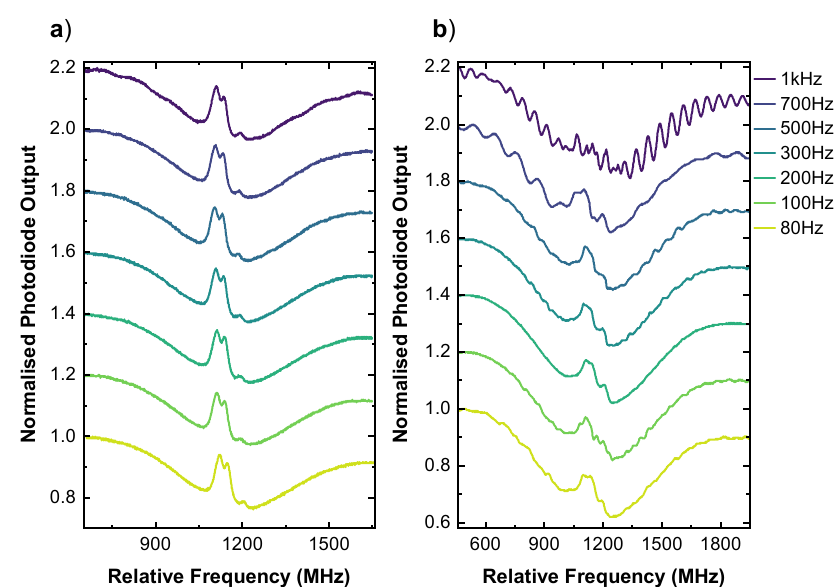} %13cm
 \caption{The effect of different ranges of single-frequency vibrations on the saturated absorption spectroscopy signal of (a) the AM optical framework directly exposed to the vibrations, and (b) a conventional setup constructed on the other side of the same optical bench. For reference, the $^{85}$Rb F=3 $\rightarrow$ F'=2,3, and 4 transition is shown. As can be seen, the compact optical system exhibits a high level of resistance to vibrational disturbance, whereas the conventional setup is extremely susceptible to the same range of oscillations. For visibility, an offset of 0.2 is applied between successive traces.}
\label{mech2}
\end{figure}

 The advantages of the 3D printed setup can be qualitatively illustrated by comparison to a conventional system. For these measurements, the aluminium plate shown in figure \ref{mechsetup} was cantilevered from the side of a standard optical bench. Data was simultaneously collected from both the compact spectroscopy apparatus on the vibrating plate and from a conventional spectroscopy apparatus located approximately two meters away on the optical bench. This apparatus employed a 130\,cm optical beam pathway from an external cavity diode laser (inside a Toptica TA pro box), through a total of six mirrors on standard kinematic mounts, terminating on a Thorlabs PD10A amplified photodiode. Note that the light was sourced from the same laser for both measurements, so disturbance of the laser cannot explain the observed differences. Fig.\,\ref{mech2} shows a comparison of the performance of the two systems. 
Fig.\,\ref{mech2}(a) shows that the form of the Doppler-free spectroscopy signal obtained from our compact spectroscopy device, mounted directly onto the vibrating plate, was barely affected by the presence of the vibrations. The same vibrations can be seen to have a substantial and deleterious effect on the spectroscopic signal produced by the conventional apparatus, despite its lesser degree of proximity to the source of the vibrations --- see Fig.\,\ref{mech2}(b). The range of frequencies applied were between 80Hz and 1kHz with an amplitude of $\pm\,0.2$\,ms$^{-2}$.
Note that this measurement demonstrates that the compact system can be operated while the vibrations are applied without major disturbance.
This clearly shows the advantages of compact AM-design, and additively manufactured frame, minimisation of the use of adjustable components and fully-surrounding support structures for the mechanical stabilisation of optical systems.

\section{Conclusions and outlook}
We have demonstrated a compact, inexpensive experimental system, based on 3D-printing methods, for performing Doppler-free spectroscopy on two lasers simultaneously, as well as generating an optical beat note with a third laser. Its response to vibrations and temperature changes has been analysed. We have thus demonstrated that the use of 3D-printing methods can lead to robust systems which can be tailored in performance to a required task.
The demonstrated device shows comparable robustness against temperature changes and vibrations to specialised, integrated systems, but at a much lower cost. It is able to operate consistently over a 28 degree temperature range and in the presence of such vibrations as might be found in an un-isolated environment, demonstrating its suitability for use in portable quantum technologies. These methods are transferable and through the application of similar techniques to other complex optical systems are likely to transform current experimental hardware. 

The system as shown is suitable for laser stabilisation for magneto-optical trapping, as needed for cold-atom-based quantum technologies. The use of the AM mount in this context allows an advanced design, tailored to the task, and can deliver high stability and reduced size and weight - and is thus relevant for the development of portable quantum technologies or the use of quantum technologies in space. 
%We have demonstrated the stability performance for a standard, inexpensive printing material (PLA). 
In the future, the capacities of 3D printing can be exploited further by tailoring stiffness and thermo-mechanical response following optimisation in a computer simulation model e.g. via latticing, structural reinforcement or material combinations. We believe that our approach of using AM materials is critical for quantum technologies and applicable to all applications requiring fixed optical configurations.
\textbf{Funding}
This work was supported by the EPSRC grants EP/R024111/1, EP/T001046/1 and EP/M013294/1 and by the European Commission grant ErBeStA (no. 800942). 

\textbf{Acknowledgment}
We would like to thank Sarah Everton and Dominic Sims for useful comments. 

\textbf{Disclosures}
The authors declare the following competing interests: N.C., L.C. and L.H. are inventors on a UK pending patent application GB 1916446.6 (applicant: University of Nottingham, inventors: Nathan Cooper, Lucia Hackerm\"{u}ller, Laurence Coles) for the device described in the article.

\textbf{Data Availability}
All data necessary to support the conclusions of the article are presented in the article. Any additional data related to this paper may be obtained from the authors upon reasonable request.

\bibliography{csa1.bib}% Produces the bibliography via BibTeX.

\begin{thebibliography}{45}%
\makeatletter
\providecommand \@ifxundefined [1]{%
 \@ifx{#1\undefined}
}%
\providecommand \@ifnum [1]{%
 \ifnum #1\expandafter \@firstoftwo
 \else \expandafter \@secondoftwo
 \fi
}%
\providecommand \@ifx [1]{%
 \ifx #1\expandafter \@firstoftwo
 \else \expandafter \@secondoftwo
 \fi
}%
\providecommand \natexlab [1]{#1}%
\providecommand \enquote  [1]{``#1''}%
\providecommand \bibnamefont  [1]{#1}%
\providecommand \bibfnamefont [1]{#1}%
\providecommand \citenamefont [1]{#1}%
\providecommand \href@noop [0]{\@secondoftwo}%
\providecommand \href [0]{\begingroup \@sanitize@url \@href}%
\providecommand \@href[1]{\@@startlink{#1}\@@href}%
\providecommand \@@href[1]{\endgroup#1\@@endlink}%
\providecommand \@sanitize@url [0]{\catcode `\\12\catcode `\$12\catcode
  `\&12\catcode `\#12\catcode `\^12\catcode `\_12\catcode `\%12\relax}%
\providecommand \@@startlink[1]{}%
\providecommand \@@endlink[0]{}%
\providecommand \url  [0]{\begingroup\@sanitize@url \@url }%
\providecommand \@url [1]{\endgroup\@href {#1}{\urlprefix }}%
\providecommand \urlprefix  [0]{URL }%
\providecommand \Eprint [0]{\href }%
\providecommand \doibase [0]{http://dx.doi.org/}%
\providecommand \selectlanguage [0]{\@gobble}%
\providecommand \bibinfo  [0]{\@secondoftwo}%
\providecommand \bibfield  [0]{\@secondoftwo}%
\providecommand \translation [1]{[#1]}%
\providecommand \BibitemOpen [0]{}%
\providecommand \bibitemStop [0]{}%
\providecommand \bibitemNoStop [0]{.\EOS\space}%
\providecommand \EOS [0]{\spacefactor3000\relax}%
\providecommand \BibitemShut  [1]{\csname bibitem#1\endcsname}%
\let\auto@bib@innerbib\@empty
%</preamble>
\bibitem [{\citenamefont {Yver}\ \emph {et~al.}(2003)\citenamefont {Yver},
  \citenamefont {Landragin}, \citenamefont {Dimarcq}, \citenamefont {Clairon},
  \citenamefont {Holleville}, \citenamefont {Cheinet}, \citenamefont {Bouyer},
  \citenamefont {Salomon},\ and\ \citenamefont {Borde}}]{YverF2003Acai}%
  \BibitemOpen
  \bibfield  {author} {\bibinfo {author} {\bibfnamefont {F.}~\bibnamefont
  {Yver}}, \bibinfo {author} {\bibfnamefont {A.}~\bibnamefont {Landragin}},
  \bibinfo {author} {\bibfnamefont {N.}~\bibnamefont {Dimarcq}}, \bibinfo
  {author} {\bibfnamefont {A.}~\bibnamefont {Clairon}}, \bibinfo {author}
  {\bibfnamefont {D.}~\bibnamefont {Holleville}}, \bibinfo {author}
  {\bibfnamefont {P.}~\bibnamefont {Cheinet}}, \bibinfo {author} {\bibfnamefont
  {P.}~\bibnamefont {Bouyer}}, \bibinfo {author} {\bibfnamefont
  {C.}~\bibnamefont {Salomon}}, \ and\ \bibinfo {author} {\bibfnamefont
  {C.}~\bibnamefont {Borde}},\ }in\ \href@noop {} {\emph {\bibinfo {booktitle}
  {2003 European Quantum Electronics Conference. EQEC 2003 (IEEE Cat No.
  03TH8665)}}}\ (\bibinfo  {publisher} {IEEE},\ \bibinfo {year} {2003})\ p.\
  \bibinfo {pages} {309}\BibitemShut {NoStop}%
\bibitem [{\citenamefont {Geiger}\ \emph {et~al.}(2020)\citenamefont {Geiger},
  \citenamefont {Landragin}, \citenamefont {Merlet},\ and\ \citenamefont
  {Pereira Dos~Santos}}]{geiger2020high}%
  \BibitemOpen
  \bibfield  {author} {\bibinfo {author} {\bibfnamefont {R.}~\bibnamefont
  {Geiger}}, \bibinfo {author} {\bibfnamefont {A.}~\bibnamefont {Landragin}},
  \bibinfo {author} {\bibfnamefont {S.}~\bibnamefont {Merlet}}, \ and\ \bibinfo
  {author} {\bibfnamefont {F.}~\bibnamefont {Pereira Dos~Santos}},\ }\href@noop
  {} {\bibfield  {journal} {\bibinfo  {journal} {AVS Quantum Science}\ }\textbf
  {\bibinfo {volume} {2}},\ \bibinfo {pages} {024702} (\bibinfo {year}
  {2020})}\BibitemShut {NoStop}%
\bibitem [{\citenamefont {Tino}(2021)}]{tino2021testing}%
  \BibitemOpen
  \bibfield  {author} {\bibinfo {author} {\bibfnamefont {G.~M.}\ \bibnamefont
  {Tino}},\ }\href@noop {} {\bibfield  {journal} {\bibinfo  {journal} {Quantum
  Science and Technology}\ }\textbf {\bibinfo {volume} {6}},\ \bibinfo {pages}
  {024014} (\bibinfo {year} {2021})}\BibitemShut {NoStop}%
\bibitem [{\citenamefont {Cacciapuoti}\ \emph {et~al.}(2020)\citenamefont
  {Cacciapuoti}, \citenamefont {Armano}, \citenamefont {Much}, \citenamefont
  {Sy}, \citenamefont {Helm}, \citenamefont {Hess}, \citenamefont {Kehrer},
  \citenamefont {Koller}, \citenamefont {Niedermaier}, \citenamefont {Esnault},
  \citenamefont {Massonnet}, \citenamefont {Goujon}, \citenamefont {Pittet},
  \citenamefont {Rochat}, \citenamefont {Liu}, \citenamefont {Schaefer},
  \citenamefont {Schwall}, \citenamefont {Prochazka}, \citenamefont {Schlicht},
  \citenamefont {Schreiber}, \citenamefont {Delva}, \citenamefont {Guerlin},
  \citenamefont {Laurent}, \citenamefont {Poncin-Lafitte}, \citenamefont
  {Lilley}, \citenamefont {Savalle}, \citenamefont {Wolf}, \citenamefont
  {Meynadier},\ and\ \citenamefont {Salomon}}]{cacciapuoti2020testing}%
  \BibitemOpen
  \bibfield  {author} {\bibinfo {author} {\bibfnamefont {L.}~\bibnamefont
  {Cacciapuoti}}, \bibinfo {author} {\bibfnamefont {M.}~\bibnamefont {Armano}},
  \bibinfo {author} {\bibfnamefont {R.}~\bibnamefont {Much}}, \bibinfo {author}
  {\bibfnamefont {O.}~\bibnamefont {Sy}}, \bibinfo {author} {\bibfnamefont
  {A.}~\bibnamefont {Helm}}, \bibinfo {author} {\bibfnamefont {M.~P.}\
  \bibnamefont {Hess}}, \bibinfo {author} {\bibfnamefont {J.}~\bibnamefont
  {Kehrer}}, \bibinfo {author} {\bibfnamefont {S.}~\bibnamefont {Koller}},
  \bibinfo {author} {\bibfnamefont {T.}~\bibnamefont {Niedermaier}}, \bibinfo
  {author} {\bibfnamefont {F.~X.}\ \bibnamefont {Esnault}}, \bibinfo {author}
  {\bibfnamefont {D.}~\bibnamefont {Massonnet}}, \bibinfo {author}
  {\bibfnamefont {D.}~\bibnamefont {Goujon}}, \bibinfo {author} {\bibfnamefont
  {J.}~\bibnamefont {Pittet}}, \bibinfo {author} {\bibfnamefont
  {P.}~\bibnamefont {Rochat}}, \bibinfo {author} {\bibfnamefont
  {S.}~\bibnamefont {Liu}}, \bibinfo {author} {\bibfnamefont {W.}~\bibnamefont
  {Schaefer}}, \bibinfo {author} {\bibfnamefont {T.}~\bibnamefont {Schwall}},
  \bibinfo {author} {\bibfnamefont {I.}~\bibnamefont {Prochazka}}, \bibinfo
  {author} {\bibfnamefont {A.}~\bibnamefont {Schlicht}}, \bibinfo {author}
  {\bibfnamefont {U.}~\bibnamefont {Schreiber}}, \bibinfo {author}
  {\bibfnamefont {P.}~\bibnamefont {Delva}}, \bibinfo {author} {\bibfnamefont
  {C.}~\bibnamefont {Guerlin}}, \bibinfo {author} {\bibfnamefont
  {P.}~\bibnamefont {Laurent}}, \bibinfo {author} {\bibfnamefont {C.~L.}\
  \bibnamefont {Poncin-Lafitte}}, \bibinfo {author} {\bibfnamefont
  {M.}~\bibnamefont {Lilley}}, \bibinfo {author} {\bibfnamefont
  {E.}~\bibnamefont {Savalle}}, \bibinfo {author} {\bibfnamefont
  {P.}~\bibnamefont {Wolf}}, \bibinfo {author} {\bibfnamefont {F.}~\bibnamefont
  {Meynadier}}, \ and\ \bibinfo {author} {\bibfnamefont {C.}~\bibnamefont
  {Salomon}},\ }\href@noop {} {\bibfield  {journal} {\bibinfo  {journal} {Eur.
  Phys. J. D}\ }\textbf {\bibinfo {volume} {74}},\ \bibinfo {pages} {1}
  (\bibinfo {year} {2020})}\BibitemShut {NoStop}%
\bibitem [{\citenamefont {Takamoto}\ \emph {et~al.}(2020)\citenamefont
  {Takamoto}, \citenamefont {Ushijima}, \citenamefont {Ohmae}, \citenamefont
  {Yahagi}, \citenamefont {Kokado}, \citenamefont {Shinkai},\ and\
  \citenamefont {Katori}}]{takamoto2020test}%
  \BibitemOpen
  \bibfield  {author} {\bibinfo {author} {\bibfnamefont {M.}~\bibnamefont
  {Takamoto}}, \bibinfo {author} {\bibfnamefont {I.}~\bibnamefont {Ushijima}},
  \bibinfo {author} {\bibfnamefont {N.}~\bibnamefont {Ohmae}}, \bibinfo
  {author} {\bibfnamefont {T.}~\bibnamefont {Yahagi}}, \bibinfo {author}
  {\bibfnamefont {K.}~\bibnamefont {Kokado}}, \bibinfo {author} {\bibfnamefont
  {H.}~\bibnamefont {Shinkai}}, \ and\ \bibinfo {author} {\bibfnamefont
  {H.}~\bibnamefont {Katori}},\ }\href@noop {} {\bibfield  {journal} {\bibinfo
  {journal} {Nat. Photon.}\ }\textbf {\bibinfo {volume} {14}},\ \bibinfo
  {pages} {411} (\bibinfo {year} {2020})}\BibitemShut {NoStop}%
\bibitem [{\citenamefont {Kaiser}\ \emph {et~al.}(2019)\citenamefont {Kaiser},
  \citenamefont {Vergyris}, \citenamefont {Martin}, \citenamefont {Aktas},
  \citenamefont {Micheli}, \citenamefont {Alibart},\ and\ \citenamefont
  {Tanzilli}}]{kaiser2019}%
  \BibitemOpen
  \bibfield  {author} {\bibinfo {author} {\bibfnamefont {F.}~\bibnamefont
  {Kaiser}}, \bibinfo {author} {\bibfnamefont {P.}~\bibnamefont {Vergyris}},
  \bibinfo {author} {\bibfnamefont {A.}~\bibnamefont {Martin}}, \bibinfo
  {author} {\bibfnamefont {D.}~\bibnamefont {Aktas}}, \bibinfo {author}
  {\bibfnamefont {M.~P.~D.}\ \bibnamefont {Micheli}}, \bibinfo {author}
  {\bibfnamefont {O.}~\bibnamefont {Alibart}}, \ and\ \bibinfo {author}
  {\bibfnamefont {S.}~\bibnamefont {Tanzilli}},\ }\href {\doibase
  10.1364/OE.27.025603} {\bibfield  {journal} {\bibinfo  {journal} {Opt.
  Express}\ }\textbf {\bibinfo {volume} {27}},\ \bibinfo {pages} {25603}
  (\bibinfo {year} {2019})}\BibitemShut {NoStop}%
\bibitem [{\citenamefont {Hague}\ \emph {et~al.}(2014)\citenamefont {Hague},
  \citenamefont {Downes}, \citenamefont {MacCormick},\ and\ \citenamefont
  {Kornilovitch}}]{HagueJP2014CRAf}%
  \BibitemOpen
  \bibfield  {author} {\bibinfo {author} {\bibfnamefont {J.~P.}\ \bibnamefont
  {Hague}}, \bibinfo {author} {\bibfnamefont {S.}~\bibnamefont {Downes}},
  \bibinfo {author} {\bibfnamefont {C.}~\bibnamefont {MacCormick}}, \ and\
  \bibinfo {author} {\bibfnamefont {P.~E.}\ \bibnamefont {Kornilovitch}},\
  }\href@noop {} {\bibfield  {journal} {\bibinfo  {journal} {J. Supercond. Nov.
  Magn.}\ }\textbf {\bibinfo {volume} {27}},\ \bibinfo {pages} {937} (\bibinfo
  {year} {2014})}\BibitemShut {NoStop}%
\bibitem [{\citenamefont {Lewenstein}\ \emph {et~al.}(2007)\citenamefont
  {Lewenstein}, \citenamefont {Sanpera}, \citenamefont {Ahufinger},
  \citenamefont {Damski}, \citenamefont {Sen},\ and\ \citenamefont
  {Sen}}]{lewenstein2007ultracold}%
  \BibitemOpen
  \bibfield  {author} {\bibinfo {author} {\bibfnamefont {M.}~\bibnamefont
  {Lewenstein}}, \bibinfo {author} {\bibfnamefont {A.}~\bibnamefont {Sanpera}},
  \bibinfo {author} {\bibfnamefont {V.}~\bibnamefont {Ahufinger}}, \bibinfo
  {author} {\bibfnamefont {B.}~\bibnamefont {Damski}}, \bibinfo {author}
  {\bibfnamefont {A.}~\bibnamefont {Sen}}, \ and\ \bibinfo {author}
  {\bibfnamefont {U.}~\bibnamefont {Sen}},\ }\href@noop {} {\bibfield
  {journal} {\bibinfo  {journal} {Advances in Physics}\ }\textbf {\bibinfo
  {volume} {56}},\ \bibinfo {pages} {243} (\bibinfo {year} {2007})}\BibitemShut
  {NoStop}%
\bibitem [{\citenamefont {Demler}\ and\ \citenamefont
  {Kitagawa}(2014)}]{demler2014strongly}%
  \BibitemOpen
  \bibfield  {author} {\bibinfo {author} {\bibfnamefont {E.}~\bibnamefont
  {Demler}}\ and\ \bibinfo {author} {\bibfnamefont {T.}~\bibnamefont
  {Kitagawa}},\ }\href@noop {} {\bibfield  {journal} {\bibinfo  {journal}
  {Lect. Notes Phys.}\ }\textbf {\bibinfo {volume} {284}},\ \bibinfo {pages}
  {28} (\bibinfo {year} {2014})}\BibitemShut {NoStop}%
\bibitem [{\citenamefont {Dalfovo}\ \emph {et~al.}(1999)\citenamefont
  {Dalfovo}, \citenamefont {Giorgini}, \citenamefont {Pitaevskii},\ and\
  \citenamefont {Stringari}}]{DalfovoFranco1999ToBc}%
  \BibitemOpen
  \bibfield  {author} {\bibinfo {author} {\bibfnamefont {F.}~\bibnamefont
  {Dalfovo}}, \bibinfo {author} {\bibfnamefont {S.}~\bibnamefont {Giorgini}},
  \bibinfo {author} {\bibfnamefont {L.~P.}\ \bibnamefont {Pitaevskii}}, \ and\
  \bibinfo {author} {\bibfnamefont {S.}~\bibnamefont {Stringari}},\ }\href@noop
  {} {\bibfield  {journal} {\bibinfo  {journal} {Rev. Mod. Phys.}\ }\textbf
  {\bibinfo {volume} {71}},\ \bibinfo {pages} {463} (\bibinfo {year}
  {1999})}\BibitemShut {NoStop}%
\bibitem [{\citenamefont {Leggett}(2001)}]{LeggettAnthonyJ2001Bcit}%
  \BibitemOpen
  \bibfield  {author} {\bibinfo {author} {\bibfnamefont {A.~J.}\ \bibnamefont
  {Leggett}},\ }\href@noop {} {\bibfield  {journal} {\bibinfo  {journal} {Rev.
  Mod. phys.}\ }\textbf {\bibinfo {volume} {73}},\ \bibinfo {pages} {307}
  (\bibinfo {year} {2001})}\BibitemShut {NoStop}%
\bibitem [{\citenamefont {Bloch}\ \emph {et~al.}(2008)\citenamefont {Bloch},
  \citenamefont {Dalibard},\ and\ \citenamefont
  {Zwerger}}]{BlochImmanuel2008Mpwu}%
  \BibitemOpen
  \bibfield  {author} {\bibinfo {author} {\bibfnamefont {I.}~\bibnamefont
  {Bloch}}, \bibinfo {author} {\bibfnamefont {J.}~\bibnamefont {Dalibard}}, \
  and\ \bibinfo {author} {\bibfnamefont {W.}~\bibnamefont {Zwerger}},\
  }\href@noop {} {\bibfield  {journal} {\bibinfo  {journal} {Rev. Mod. Phys.}\
  }\textbf {\bibinfo {volume} {80}},\ \bibinfo {pages} {885} (\bibinfo {year}
  {2008})}\BibitemShut {NoStop}%
\bibitem [{\citenamefont {Giorgini}\ \emph {et~al.}(2008)\citenamefont
  {Giorgini}, \citenamefont {Pitaevskii},\ and\ \citenamefont
  {Stringari}}]{GiorginiStefano2008Toua}%
  \BibitemOpen
  \bibfield  {author} {\bibinfo {author} {\bibfnamefont {S.}~\bibnamefont
  {Giorgini}}, \bibinfo {author} {\bibfnamefont {L.~P.}\ \bibnamefont
  {Pitaevskii}}, \ and\ \bibinfo {author} {\bibfnamefont {S.}~\bibnamefont
  {Stringari}},\ }\href@noop {} {\bibfield  {journal} {\bibinfo  {journal}
  {Rev. Mod. Phys.}\ }\textbf {\bibinfo {volume} {80}},\ \bibinfo {pages}
  {1215} (\bibinfo {year} {2008})}\BibitemShut {NoStop}%
\bibitem [{\citenamefont {Farrera}\ \emph {et~al.}(2016)\citenamefont
  {Farrera}, \citenamefont {Heinze}, \citenamefont {Albrecht}, \citenamefont
  {Ho}, \citenamefont {Ch{\'a}vez}, \citenamefont {Teo}, \citenamefont
  {Sangouard},\ and\ \citenamefont {De~Riedmatten}}]{farrera2016generation}%
  \BibitemOpen
  \bibfield  {author} {\bibinfo {author} {\bibfnamefont {P.}~\bibnamefont
  {Farrera}}, \bibinfo {author} {\bibfnamefont {G.}~\bibnamefont {Heinze}},
  \bibinfo {author} {\bibfnamefont {B.}~\bibnamefont {Albrecht}}, \bibinfo
  {author} {\bibfnamefont {M.}~\bibnamefont {Ho}}, \bibinfo {author}
  {\bibfnamefont {M.}~\bibnamefont {Ch{\'a}vez}}, \bibinfo {author}
  {\bibfnamefont {C.}~\bibnamefont {Teo}}, \bibinfo {author} {\bibfnamefont
  {N.}~\bibnamefont {Sangouard}}, \ and\ \bibinfo {author} {\bibfnamefont
  {H.}~\bibnamefont {De~Riedmatten}},\ }\href@noop {} {\bibfield  {journal}
  {\bibinfo  {journal} {Nat. Commun.}\ }\textbf {\bibinfo {volume} {7}},\
  \bibinfo {pages} {1} (\bibinfo {year} {2016})}\BibitemShut {NoStop}%
\bibitem [{\citenamefont {Garcia-Ripoll}\ \emph {et~al.}(2005)\citenamefont
  {Garcia-Ripoll}, \citenamefont {Zoller},\ and\ \citenamefont
  {Cirac}}]{Garcia-RipollJJ2005Qipw}%
  \BibitemOpen
  \bibfield  {author} {\bibinfo {author} {\bibfnamefont {J.~J.}\ \bibnamefont
  {Garcia-Ripoll}}, \bibinfo {author} {\bibfnamefont {P.}~\bibnamefont
  {Zoller}}, \ and\ \bibinfo {author} {\bibfnamefont {J.~I.}\ \bibnamefont
  {Cirac}},\ }\href@noop {} {\bibfield  {journal} {\bibinfo  {journal} {J.
  Phys. B At. Mol. Opt. Phys.}\ }\textbf {\bibinfo {volume} {38}},\ \bibinfo
  {pages} {S567} (\bibinfo {year} {2005})}\BibitemShut {NoStop}%
\bibitem [{\citenamefont {Ryabtsev}\ \emph {et~al.}(2016)\citenamefont
  {Ryabtsev}, \citenamefont {Beterov}, \citenamefont {Tretyakov}, \citenamefont
  {Entin},\ and\ \citenamefont {Yakshina}}]{RyabtsevII2016Socr}%
  \BibitemOpen
  \bibfield  {author} {\bibinfo {author} {\bibfnamefont {I.~I.}\ \bibnamefont
  {Ryabtsev}}, \bibinfo {author} {\bibfnamefont {I.~I.}\ \bibnamefont
  {Beterov}}, \bibinfo {author} {\bibfnamefont {D.~B.}\ \bibnamefont
  {Tretyakov}}, \bibinfo {author} {\bibfnamefont {V.~M.}\ \bibnamefont
  {Entin}}, \ and\ \bibinfo {author} {\bibfnamefont {E.~A.}\ \bibnamefont
  {Yakshina}},\ }\href@noop {} {\bibfield  {journal} {\bibinfo  {journal}
  {Phys.-Uspekhi}\ }\textbf {\bibinfo {volume} {59}},\ \bibinfo {pages} {196}
  (\bibinfo {year} {2016})}\BibitemShut {NoStop}%
\bibitem [{\citenamefont {Roztocki}\ \emph {et~al.}(2017)\citenamefont
  {Roztocki}, \citenamefont {Kues}, \citenamefont {Reimer}, \citenamefont
  {Wetzel}, \citenamefont {Sciara}, \citenamefont {Zhang}, \citenamefont
  {Cino}, \citenamefont {Little}, \citenamefont {Chu}, \citenamefont {Moss},\
  and\ \citenamefont {Morandotti}}]{roztocki2017}%
  \BibitemOpen
  \bibfield  {author} {\bibinfo {author} {\bibfnamefont {P.}~\bibnamefont
  {Roztocki}}, \bibinfo {author} {\bibfnamefont {M.}~\bibnamefont {Kues}},
  \bibinfo {author} {\bibfnamefont {C.}~\bibnamefont {Reimer}}, \bibinfo
  {author} {\bibfnamefont {B.}~\bibnamefont {Wetzel}}, \bibinfo {author}
  {\bibfnamefont {S.}~\bibnamefont {Sciara}}, \bibinfo {author} {\bibfnamefont
  {Y.}~\bibnamefont {Zhang}}, \bibinfo {author} {\bibfnamefont
  {A.}~\bibnamefont {Cino}}, \bibinfo {author} {\bibfnamefont {B.~E.}\
  \bibnamefont {Little}}, \bibinfo {author} {\bibfnamefont {S.~T.}\
  \bibnamefont {Chu}}, \bibinfo {author} {\bibfnamefont {D.~J.}\ \bibnamefont
  {Moss}}, \ and\ \bibinfo {author} {\bibfnamefont {R.}~\bibnamefont
  {Morandotti}},\ }\href {\doibase 10.1364/OE.25.018940} {\bibfield  {journal}
  {\bibinfo  {journal} {Opt. Express}\ }\textbf {\bibinfo {volume} {25}},\
  \bibinfo {pages} {18940} (\bibinfo {year} {2017})}\BibitemShut {NoStop}%
\bibitem [{\citenamefont {Schneeweiss}\ \emph {et~al.}(2018)\citenamefont
  {Schneeweiss}, \citenamefont {Dareau},\ and\ \citenamefont
  {Sayrin}}]{schneeweiss2018cold}%
  \BibitemOpen
  \bibfield  {author} {\bibinfo {author} {\bibfnamefont {P.}~\bibnamefont
  {Schneeweiss}}, \bibinfo {author} {\bibfnamefont {A.}~\bibnamefont {Dareau}},
  \ and\ \bibinfo {author} {\bibfnamefont {C.}~\bibnamefont {Sayrin}},\
  }\href@noop {} {\bibfield  {journal} {\bibinfo  {journal} {Physical Review
  A}\ }\textbf {\bibinfo {volume} {98}},\ \bibinfo {pages} {021801} (\bibinfo
  {year} {2018})}\BibitemShut {NoStop}%
\bibitem [{\citenamefont {Kuzmich}\ \emph {et~al.}(2003)\citenamefont
  {Kuzmich}, \citenamefont {Bowen}, \citenamefont {Boozer}, \citenamefont
  {Boca}, \citenamefont {Chou}, \citenamefont {Duan},\ and\ \citenamefont
  {Kimble}}]{kuzmich2003generation}%
  \BibitemOpen
  \bibfield  {author} {\bibinfo {author} {\bibfnamefont {A.}~\bibnamefont
  {Kuzmich}}, \bibinfo {author} {\bibfnamefont {W.}~\bibnamefont {Bowen}},
  \bibinfo {author} {\bibfnamefont {A.}~\bibnamefont {Boozer}}, \bibinfo
  {author} {\bibfnamefont {A.}~\bibnamefont {Boca}}, \bibinfo {author}
  {\bibfnamefont {C.}~\bibnamefont {Chou}}, \bibinfo {author} {\bibfnamefont
  {L.-M.}\ \bibnamefont {Duan}}, \ and\ \bibinfo {author} {\bibfnamefont
  {H.}~\bibnamefont {Kimble}},\ }\href@noop {} {\bibfield  {journal} {\bibinfo
  {journal} {Nature}\ }\textbf {\bibinfo {volume} {423}},\ \bibinfo {pages}
  {731} (\bibinfo {year} {2003})}\BibitemShut {NoStop}%
\bibitem [{\citenamefont {Bao}\ \emph {et~al.}(2012)\citenamefont {Bao},
  \citenamefont {Reingruber}, \citenamefont {Dietrich}, \citenamefont {Rui},
  \citenamefont {D{\"u}ck}, \citenamefont {Strassel}, \citenamefont {Li},
  \citenamefont {Liu}, \citenamefont {Zhao},\ and\ \citenamefont
  {Pan}}]{bao2012efficient}%
  \BibitemOpen
  \bibfield  {author} {\bibinfo {author} {\bibfnamefont {X.-H.}\ \bibnamefont
  {Bao}}, \bibinfo {author} {\bibfnamefont {A.}~\bibnamefont {Reingruber}},
  \bibinfo {author} {\bibfnamefont {P.}~\bibnamefont {Dietrich}}, \bibinfo
  {author} {\bibfnamefont {J.}~\bibnamefont {Rui}}, \bibinfo {author}
  {\bibfnamefont {A.}~\bibnamefont {D{\"u}ck}}, \bibinfo {author}
  {\bibfnamefont {T.}~\bibnamefont {Strassel}}, \bibinfo {author}
  {\bibfnamefont {L.}~\bibnamefont {Li}}, \bibinfo {author} {\bibfnamefont
  {N.-L.}\ \bibnamefont {Liu}}, \bibinfo {author} {\bibfnamefont
  {B.}~\bibnamefont {Zhao}}, \ and\ \bibinfo {author} {\bibfnamefont {J.-W.}\
  \bibnamefont {Pan}},\ }\href@noop {} {\bibfield  {journal} {\bibinfo
  {journal} {Nature Physics}\ }\textbf {\bibinfo {volume} {8}},\ \bibinfo
  {pages} {517} (\bibinfo {year} {2012})}\BibitemShut {NoStop}%
\bibitem [{\citenamefont {Bongs}\ \emph {et~al.}(2014)\citenamefont {Bongs},
  \citenamefont {Malcolm}, \citenamefont {Ramelloo}, \citenamefont {Zhu},
  \citenamefont {Boyer}, \citenamefont {Valenzuela}, \citenamefont {Maclean},
  \citenamefont {Piccardo-Selg}, \citenamefont {Mellor}, \citenamefont
  {Fernholz}, \citenamefont {Fromhold}, \citenamefont {Krüger}, \citenamefont
  {Hellmig}, \citenamefont {Grote}, \citenamefont {Dörscher}, \citenamefont
  {Duncker}, \citenamefont {Windpassinger}, \citenamefont {Sengstock},
  \citenamefont {Becker}, \citenamefont {Pelle}, \citenamefont {Hilico},
  \citenamefont {Zhou}, \citenamefont {Angonin}, \citenamefont {Wolf},
  \citenamefont {Dos~Santos}, \citenamefont {Bertoldi}, \citenamefont {Bouyer},
  \citenamefont {Mazzoni}, \citenamefont {Poli}, \citenamefont {Sorrentino},
  \citenamefont {Tarallo}, \citenamefont {Tino}, \citenamefont {Stellmer},
  \citenamefont {Schreck}, \citenamefont {Popp}, \citenamefont {Herr},
  \citenamefont {Wendrich}, \citenamefont {Ertmer}, \citenamefont {Rasel},
  \citenamefont {Kürbis},\ and\ \citenamefont {Wicht}}]{bongs2014isense}%
  \BibitemOpen
  \bibfield  {author} {\bibinfo {author} {\bibfnamefont {K.}~\bibnamefont
  {Bongs}}, \bibinfo {author} {\bibfnamefont {J.}~\bibnamefont {Malcolm}},
  \bibinfo {author} {\bibfnamefont {C.}~\bibnamefont {Ramelloo}}, \bibinfo
  {author} {\bibfnamefont {L.}~\bibnamefont {Zhu}}, \bibinfo {author}
  {\bibfnamefont {V.}~\bibnamefont {Boyer}}, \bibinfo {author} {\bibfnamefont
  {T.}~\bibnamefont {Valenzuela}}, \bibinfo {author} {\bibfnamefont
  {J.}~\bibnamefont {Maclean}}, \bibinfo {author} {\bibfnamefont
  {A.}~\bibnamefont {Piccardo-Selg}}, \bibinfo {author} {\bibfnamefont
  {C.}~\bibnamefont {Mellor}}, \bibinfo {author} {\bibfnamefont
  {T.}~\bibnamefont {Fernholz}}, \bibinfo {author} {\bibfnamefont
  {M.}~\bibnamefont {Fromhold}}, \bibinfo {author} {\bibfnamefont
  {P.}~\bibnamefont {Krüger}}, \bibinfo {author} {\bibfnamefont
  {O.}~\bibnamefont {Hellmig}}, \bibinfo {author} {\bibfnamefont
  {A.}~\bibnamefont {Grote}}, \bibinfo {author} {\bibfnamefont
  {S.}~\bibnamefont {Dörscher}}, \bibinfo {author} {\bibfnamefont
  {H.}~\bibnamefont {Duncker}}, \bibinfo {author} {\bibfnamefont
  {P.}~\bibnamefont {Windpassinger}}, \bibinfo {author} {\bibfnamefont
  {K.}~\bibnamefont {Sengstock}}, \bibinfo {author} {\bibfnamefont
  {C.}~\bibnamefont {Becker}}, \bibinfo {author} {\bibfnamefont
  {B.}~\bibnamefont {Pelle}}, \bibinfo {author} {\bibfnamefont
  {A.}~\bibnamefont {Hilico}}, \bibinfo {author} {\bibfnamefont
  {M.}~\bibnamefont {Zhou}}, \bibinfo {author} {\bibfnamefont {M.-C.}\
  \bibnamefont {Angonin}}, \bibinfo {author} {\bibfnamefont {P.}~\bibnamefont
  {Wolf}}, \bibinfo {author} {\bibfnamefont {F.~P.}\ \bibnamefont
  {Dos~Santos}}, \bibinfo {author} {\bibfnamefont {F.}~\bibnamefont
  {Bertoldi}}, \bibinfo {author} {\bibfnamefont {P.}~\bibnamefont {Bouyer}},
  \bibinfo {author} {\bibfnamefont {T.}~\bibnamefont {Mazzoni}}, \bibinfo
  {author} {\bibfnamefont {N.}~\bibnamefont {Poli}}, \bibinfo {author}
  {\bibfnamefont {F.}~\bibnamefont {Sorrentino}}, \bibinfo {author}
  {\bibfnamefont {M.}~\bibnamefont {Tarallo}}, \bibinfo {author} {\bibfnamefont
  {G.}~\bibnamefont {Tino}}, \bibinfo {author} {\bibfnamefont {S.}~\bibnamefont
  {Stellmer}}, \bibinfo {author} {\bibfnamefont {F.}~\bibnamefont {Schreck}},
  \bibinfo {author} {\bibfnamefont {M.}~\bibnamefont {Popp}}, \bibinfo {author}
  {\bibfnamefont {W.}~\bibnamefont {Herr}}, \bibinfo {author} {\bibfnamefont
  {T.}~\bibnamefont {Wendrich}}, \bibinfo {author} {\bibfnamefont
  {W.}~\bibnamefont {Ertmer}}, \bibinfo {author} {\bibfnamefont
  {E.}~\bibnamefont {Rasel}}, \bibinfo {author} {\bibfnamefont
  {A.}~\bibnamefont {Kürbis}, \bibfnamefont {Christian~andPeters}}, \ and\
  \bibinfo {author} {\bibfnamefont {A.}~\bibnamefont {Wicht}},\ }in\ \href@noop
  {} {\emph {\bibinfo {booktitle} {Quantum Information and Measurement}}}\
  (\bibinfo {organization} {Optical Society of America},\ \bibinfo {year}
  {2014})\ pp.\ \bibinfo {pages} {QTu3B--1}\BibitemShut {NoStop}%
\bibitem [{\citenamefont {Falke}\ \emph {et~al.}(2014)\citenamefont {Falke},
  \citenamefont {Lemke}, \citenamefont {Grebing}, \citenamefont {Lipphardt},
  \citenamefont {Weyers}, \citenamefont {Gerginov}, \citenamefont {Huntemann},
  \citenamefont {Hagemann}, \citenamefont {Al-Masoudi}, \citenamefont
  {Häfner}, \citenamefont {Vogt}, \citenamefont {Sterr}, ,\ and\ \citenamefont
  {Lisdat}}]{falke2014strontium}%
  \BibitemOpen
  \bibfield  {author} {\bibinfo {author} {\bibfnamefont {S.}~\bibnamefont
  {Falke}}, \bibinfo {author} {\bibfnamefont {N.}~\bibnamefont {Lemke}},
  \bibinfo {author} {\bibfnamefont {C.}~\bibnamefont {Grebing}}, \bibinfo
  {author} {\bibfnamefont {B.}~\bibnamefont {Lipphardt}}, \bibinfo {author}
  {\bibfnamefont {S.}~\bibnamefont {Weyers}}, \bibinfo {author} {\bibfnamefont
  {V.}~\bibnamefont {Gerginov}}, \bibinfo {author} {\bibfnamefont
  {N.}~\bibnamefont {Huntemann}}, \bibinfo {author} {\bibfnamefont
  {C.}~\bibnamefont {Hagemann}}, \bibinfo {author} {\bibfnamefont
  {A.}~\bibnamefont {Al-Masoudi}}, \bibinfo {author} {\bibfnamefont
  {S.}~\bibnamefont {Häfner}}, \bibinfo {author} {\bibfnamefont
  {S.}~\bibnamefont {Vogt}}, \bibinfo {author} {\bibfnamefont {U.}~\bibnamefont
  {Sterr}}, , \ and\ \bibinfo {author} {\bibfnamefont {C.}~\bibnamefont
  {Lisdat}},\ }\href@noop {} {\bibfield  {journal} {\bibinfo  {journal} {New
  Journal of Physics}\ }\textbf {\bibinfo {volume} {16}},\ \bibinfo {pages}
  {073023} (\bibinfo {year} {2014})}\BibitemShut {NoStop}%
\bibitem [{\citenamefont {Ludlow}\ \emph {et~al.}(2015)\citenamefont {Ludlow},
  \citenamefont {Boyd}, \citenamefont {Ye}, \citenamefont {Peik},\ and\
  \citenamefont {Schmidt}}]{ludlow2015optical}%
  \BibitemOpen
  \bibfield  {author} {\bibinfo {author} {\bibfnamefont {A.~D.}\ \bibnamefont
  {Ludlow}}, \bibinfo {author} {\bibfnamefont {M.~M.}\ \bibnamefont {Boyd}},
  \bibinfo {author} {\bibfnamefont {J.}~\bibnamefont {Ye}}, \bibinfo {author}
  {\bibfnamefont {E.}~\bibnamefont {Peik}}, \ and\ \bibinfo {author}
  {\bibfnamefont {P.~O.}\ \bibnamefont {Schmidt}},\ }\href@noop {} {\bibfield
  {journal} {\bibinfo  {journal} {Rev. Mod. Phys.}\ }\textbf {\bibinfo {volume}
  {87}},\ \bibinfo {pages} {637} (\bibinfo {year} {2015})}\BibitemShut
  {NoStop}%
\bibitem [{\citenamefont {Riehle}(2017)}]{riehle2017optical}%
  \BibitemOpen
  \bibfield  {author} {\bibinfo {author} {\bibfnamefont {F.}~\bibnamefont
  {Riehle}},\ }\href@noop {} {\bibfield  {journal} {\bibinfo  {journal} {Nat.
  Photon.}\ }\textbf {\bibinfo {volume} {11}},\ \bibinfo {pages} {25} (\bibinfo
  {year} {2017})}\BibitemShut {NoStop}%
\bibitem [{\citenamefont {Kemp}\ \emph {et~al.}(2016)\citenamefont {Kemp},
  \citenamefont {Butler}, \citenamefont {Freytag}, \citenamefont {Hopkins},
  \citenamefont {Hinds}, \citenamefont {Tarbutt},\ and\ \citenamefont
  {Cornish}}]{zeeman1}%
  \BibitemOpen
  \bibfield  {author} {\bibinfo {author} {\bibfnamefont {S.~L.}\ \bibnamefont
  {Kemp}}, \bibinfo {author} {\bibfnamefont {K.~L.}\ \bibnamefont {Butler}},
  \bibinfo {author} {\bibfnamefont {R.}~\bibnamefont {Freytag}}, \bibinfo
  {author} {\bibfnamefont {S.~A.}\ \bibnamefont {Hopkins}}, \bibinfo {author}
  {\bibfnamefont {E.~A.}\ \bibnamefont {Hinds}}, \bibinfo {author}
  {\bibfnamefont {M.~R.}\ \bibnamefont {Tarbutt}}, \ and\ \bibinfo {author}
  {\bibfnamefont {S.~L.}\ \bibnamefont {Cornish}},\ }\href {\doibase
  10.1063/1.4941719} {\bibfield  {journal} {\bibinfo  {journal} {Rev. Sci.
  Instrum.}\ }\textbf {\bibinfo {volume} {87}},\ \bibinfo {pages} {023105+}
  (\bibinfo {year} {2016})}\BibitemShut {NoStop}%
\bibitem [{\citenamefont {Cooper}\ \emph
  {et~al.}(2021{\natexlab{a}})\citenamefont {Cooper}, \citenamefont {Coles},
  \citenamefont {Everton}, \citenamefont {Maskery}, \citenamefont {Campion},
  \citenamefont {Madkhaly}, \citenamefont {Morley}, \citenamefont {O’Shea},
  \citenamefont {Evans}, \citenamefont {Saint}, \citenamefont {Krüger},
  \citenamefont {Oručević}, \citenamefont {Tuck}, \citenamefont {Wildman},
  \citenamefont {Fromhold},\ and\ \citenamefont {Hackermüller}}]{Cooper2021}%
  \BibitemOpen
  \bibfield  {author} {\bibinfo {author} {\bibfnamefont {N.}~\bibnamefont
  {Cooper}}, \bibinfo {author} {\bibfnamefont {L.}~\bibnamefont {Coles}},
  \bibinfo {author} {\bibfnamefont {S.}~\bibnamefont {Everton}}, \bibinfo
  {author} {\bibfnamefont {I.}~\bibnamefont {Maskery}}, \bibinfo {author}
  {\bibfnamefont {R.}~\bibnamefont {Campion}}, \bibinfo {author} {\bibfnamefont
  {S.}~\bibnamefont {Madkhaly}}, \bibinfo {author} {\bibfnamefont
  {C.}~\bibnamefont {Morley}}, \bibinfo {author} {\bibfnamefont
  {J.}~\bibnamefont {O’Shea}}, \bibinfo {author} {\bibfnamefont
  {W.}~\bibnamefont {Evans}}, \bibinfo {author} {\bibfnamefont
  {R.}~\bibnamefont {Saint}}, \bibinfo {author} {\bibfnamefont
  {P.}~\bibnamefont {Krüger}}, \bibinfo {author} {\bibfnamefont
  {F.}~\bibnamefont {Oručević}}, \bibinfo {author} {\bibfnamefont
  {C.}~\bibnamefont {Tuck}}, \bibinfo {author} {\bibfnamefont {R.}~\bibnamefont
  {Wildman}}, \bibinfo {author} {\bibfnamefont {T.}~\bibnamefont {Fromhold}}, \
  and\ \bibinfo {author} {\bibfnamefont {L.}~\bibnamefont {Hackermüller}},\
  }\href@noop {} {\bibfield  {journal} {\bibinfo  {journal} {Addit. Manuf.}\
  }\textbf {\bibinfo {volume} {40}},\ \bibinfo {pages} {101898} (\bibinfo
  {year} {2021}{\natexlab{a}})}\BibitemShut {NoStop}%
\bibitem [{\citenamefont {Madkhaly}\ \emph {et~al.}(2021)\citenamefont
  {Madkhaly}, \citenamefont {Coles}, \citenamefont {Morley}, \citenamefont
  {Colquhoun}, \citenamefont {Fromhold}, \citenamefont {Cooper},\ and\
  \citenamefont {Hackerm\"uller}}]{optamot1}%
  \BibitemOpen
  \bibfield  {author} {\bibinfo {author} {\bibfnamefont {S.}~\bibnamefont
  {Madkhaly}}, \bibinfo {author} {\bibfnamefont {L.}~\bibnamefont {Coles}},
  \bibinfo {author} {\bibfnamefont {C.}~\bibnamefont {Morley}}, \bibinfo
  {author} {\bibfnamefont {C.}~\bibnamefont {Colquhoun}}, \bibinfo {author}
  {\bibfnamefont {T.}~\bibnamefont {Fromhold}}, \bibinfo {author}
  {\bibfnamefont {N.}~\bibnamefont {Cooper}}, \ and\ \bibinfo {author}
  {\bibfnamefont {L.}~\bibnamefont {Hackerm\"uller}},\ }\href {\doibase
  10.1103/PRXQuantum.2.030326} {\bibfield  {journal} {\bibinfo  {journal} {PRX
  Quantum}\ }\textbf {\bibinfo {volume} {2}},\ \bibinfo {pages} {030326}
  (\bibinfo {year} {2021})}\BibitemShut {NoStop}%
\bibitem [{\citenamefont {Jared}\ \emph {et~al.}(2017)\citenamefont {Jared},
  \citenamefont {Aguilo}, \citenamefont {Beghini}, \citenamefont {Boyce},
  \citenamefont {Clark}, \citenamefont {Cook}, \citenamefont {Kaehr},\ and\
  \citenamefont {Robbins}}]{jared2017additive}%
  \BibitemOpen
  \bibfield  {author} {\bibinfo {author} {\bibfnamefont {B.~H.}\ \bibnamefont
  {Jared}}, \bibinfo {author} {\bibfnamefont {M.~A.}\ \bibnamefont {Aguilo}},
  \bibinfo {author} {\bibfnamefont {L.~L.}\ \bibnamefont {Beghini}}, \bibinfo
  {author} {\bibfnamefont {B.~L.}\ \bibnamefont {Boyce}}, \bibinfo {author}
  {\bibfnamefont {B.~W.}\ \bibnamefont {Clark}}, \bibinfo {author}
  {\bibfnamefont {A.}~\bibnamefont {Cook}}, \bibinfo {author} {\bibfnamefont
  {B.~J.}\ \bibnamefont {Kaehr}}, \ and\ \bibinfo {author} {\bibfnamefont
  {J.}~\bibnamefont {Robbins}},\ }\href@noop {} {\bibfield  {journal} {\bibinfo
   {journal} {Scr. Mater}\ }\textbf {\bibinfo {volume} {135}},\ \bibinfo
  {pages} {141} (\bibinfo {year} {2017})}\BibitemShut {NoStop}%
\bibitem [{\citenamefont {McGilligan}\ \emph {et~al.}(2017)\citenamefont
  {McGilligan}, \citenamefont {Griffin}, \citenamefont {Elvin}, \citenamefont
  {Ingleby}, \citenamefont {Riis},\ and\ \citenamefont
  {Arnold}}]{mcgilligan2017grating}%
  \BibitemOpen
  \bibfield  {author} {\bibinfo {author} {\bibfnamefont {J.~P.}\ \bibnamefont
  {McGilligan}}, \bibinfo {author} {\bibfnamefont {P.~F.}\ \bibnamefont
  {Griffin}}, \bibinfo {author} {\bibfnamefont {R.}~\bibnamefont {Elvin}},
  \bibinfo {author} {\bibfnamefont {S.~J.}\ \bibnamefont {Ingleby}}, \bibinfo
  {author} {\bibfnamefont {E.}~\bibnamefont {Riis}}, \ and\ \bibinfo {author}
  {\bibfnamefont {A.~S.}\ \bibnamefont {Arnold}},\ }\href@noop {} {\bibfield
  {journal} {\bibinfo  {journal} {Sci. Rep.}\ }\textbf {\bibinfo {volume}
  {7}},\ \bibinfo {pages} {1} (\bibinfo {year} {2017})}\BibitemShut {NoStop}%
\bibitem [{\citenamefont {Schkolnik}\ \emph {et~al.}(2017)\citenamefont
  {Schkolnik}, \citenamefont {D{\"o}ringshoff}, \citenamefont {Gutsch},
  \citenamefont {Oswald}, \citenamefont {Schuldt}, \citenamefont {Braxmaier},
  \citenamefont {Lezius}, \citenamefont {Holzwarth}, \citenamefont
  {K{\"u}rbis}, \citenamefont {Bawamia}, \citenamefont {Krutzik},\ and\
  \citenamefont {Peters}}]{rocket2017jokarus}%
  \BibitemOpen
  \bibfield  {author} {\bibinfo {author} {\bibfnamefont {V.}~\bibnamefont
  {Schkolnik}}, \bibinfo {author} {\bibfnamefont {K.}~\bibnamefont
  {D{\"o}ringshoff}}, \bibinfo {author} {\bibfnamefont {F.~B.}\ \bibnamefont
  {Gutsch}}, \bibinfo {author} {\bibfnamefont {M.}~\bibnamefont {Oswald}},
  \bibinfo {author} {\bibfnamefont {T.}~\bibnamefont {Schuldt}}, \bibinfo
  {author} {\bibfnamefont {C.}~\bibnamefont {Braxmaier}}, \bibinfo {author}
  {\bibfnamefont {M.}~\bibnamefont {Lezius}}, \bibinfo {author} {\bibfnamefont
  {R.}~\bibnamefont {Holzwarth}}, \bibinfo {author} {\bibfnamefont
  {C.}~\bibnamefont {K{\"u}rbis}}, \bibinfo {author} {\bibfnamefont
  {A.}~\bibnamefont {Bawamia}}, \bibinfo {author} {\bibfnamefont
  {M.}~\bibnamefont {Krutzik}}, \ and\ \bibinfo {author} {\bibfnamefont
  {A.}~\bibnamefont {Peters}},\ }\href@noop {} {\bibfield  {journal} {\bibinfo
  {journal} {EPJ Quantum Technology}\ }\textbf {\bibinfo {volume} {4}},\
  \bibinfo {pages} {1} (\bibinfo {year} {2017})}\BibitemShut {NoStop}%
\bibitem [{\citenamefont {Strangfeld}\ \emph {et~al.}(2021)\citenamefont
  {Strangfeld}, \citenamefont {Kanthak}, \citenamefont {Schiemangk},
  \citenamefont {Wiegand}, \citenamefont {Wicht}, \citenamefont {Ling},\ and\
  \citenamefont {Krutzik}}]{strangfeld2021prototype}%
  \BibitemOpen
  \bibfield  {author} {\bibinfo {author} {\bibfnamefont {A.}~\bibnamefont
  {Strangfeld}}, \bibinfo {author} {\bibfnamefont {S.}~\bibnamefont {Kanthak}},
  \bibinfo {author} {\bibfnamefont {M.}~\bibnamefont {Schiemangk}}, \bibinfo
  {author} {\bibfnamefont {B.}~\bibnamefont {Wiegand}}, \bibinfo {author}
  {\bibfnamefont {A.}~\bibnamefont {Wicht}}, \bibinfo {author} {\bibfnamefont
  {A.}~\bibnamefont {Ling}}, \ and\ \bibinfo {author} {\bibfnamefont
  {M.}~\bibnamefont {Krutzik}},\ }\href@noop {} {\bibfield  {journal} {\bibinfo
   {journal} {J Opt Soc Am B.}\ }\textbf {\bibinfo {volume} {38}},\ \bibinfo
  {pages} {1885} (\bibinfo {year} {2021})}\BibitemShut {NoStop}%
\bibitem [{\citenamefont {Belenchia}\ \emph {et~al.}(2022)\citenamefont
  {Belenchia}, \citenamefont {Carlesso}, \citenamefont {Bayraktar},
  \citenamefont {Dequal}, \citenamefont {Derkach}, \citenamefont {Gasbarri},
  \citenamefont {Herr}, \citenamefont {Li}, \citenamefont {Rademacher},
  \citenamefont {Sidhu}, \citenamefont {Oi}, \citenamefont {Seidel},
  \citenamefont {Kaltenbaek}, \citenamefont {Marquardt}, \citenamefont
  {Ulbricht}, \citenamefont {Usenko}, \citenamefont {Wörner}, \citenamefont
  {Xuereb}, \citenamefont {Paternostro},\ and\ \citenamefont
  {Bassi}}]{belenchia2022quantum}%
  \BibitemOpen
  \bibfield  {author} {\bibinfo {author} {\bibfnamefont {A.}~\bibnamefont
  {Belenchia}}, \bibinfo {author} {\bibfnamefont {M.}~\bibnamefont {Carlesso}},
  \bibinfo {author} {\bibfnamefont {{\"O}.}~\bibnamefont {Bayraktar}}, \bibinfo
  {author} {\bibfnamefont {D.}~\bibnamefont {Dequal}}, \bibinfo {author}
  {\bibfnamefont {I.}~\bibnamefont {Derkach}}, \bibinfo {author} {\bibfnamefont
  {G.}~\bibnamefont {Gasbarri}}, \bibinfo {author} {\bibfnamefont
  {W.}~\bibnamefont {Herr}}, \bibinfo {author} {\bibfnamefont {Y.~L.}\
  \bibnamefont {Li}}, \bibinfo {author} {\bibfnamefont {M.}~\bibnamefont
  {Rademacher}}, \bibinfo {author} {\bibfnamefont {J.}~\bibnamefont {Sidhu}},
  \bibinfo {author} {\bibfnamefont {D.~K.}\ \bibnamefont {Oi}}, \bibinfo
  {author} {\bibfnamefont {S.~T.}\ \bibnamefont {Seidel}}, \bibinfo {author}
  {\bibfnamefont {R.}~\bibnamefont {Kaltenbaek}}, \bibinfo {author}
  {\bibfnamefont {C.}~\bibnamefont {Marquardt}}, \bibinfo {author}
  {\bibfnamefont {H.}~\bibnamefont {Ulbricht}}, \bibinfo {author}
  {\bibfnamefont {V.~C.}\ \bibnamefont {Usenko}}, \bibinfo {author}
  {\bibfnamefont {L.}~\bibnamefont {Wörner}}, \bibinfo {author} {\bibfnamefont
  {A.}~\bibnamefont {Xuereb}}, \bibinfo {author} {\bibfnamefont
  {M.}~\bibnamefont {Paternostro}}, \ and\ \bibinfo {author} {\bibfnamefont
  {A.}~\bibnamefont {Bassi}},\ }\href@noop {} {\bibfield  {journal} {\bibinfo
  {journal} {Phys. Rep.}\ }\textbf {\bibinfo {volume} {951}},\ \bibinfo {pages}
  {1} (\bibinfo {year} {2022})}\BibitemShut {NoStop}%
\bibitem [{\citenamefont {Frye}\ \emph {et~al.}(2021)\citenamefont {Frye},
  \citenamefont {Abend}, \citenamefont {Bartosch}, \citenamefont {Bawamia},
  \citenamefont {Becker}, \citenamefont {Blume}, \citenamefont {Braxmaier},
  \citenamefont {Chiow}, \citenamefont {Efremov}, \citenamefont {Ertmer},
  \citenamefont {Fierlinger}, \citenamefont {Franz}, \citenamefont {Gaaloul},
  \citenamefont {Grosse}, \citenamefont {Grzeschik}, \citenamefont {Hellmig},
  \citenamefont {Henderson}, \citenamefont {Herr}, \citenamefont {Israelsson},
  \citenamefont {Kohel}, \citenamefont {Krutzik}, \citenamefont {Kürbis},
  \citenamefont {Lämmerzahl}, \citenamefont {List}, \citenamefont {Lüdtke},
  \citenamefont {Lundblad}, \citenamefont {Marburger}, \citenamefont {Meister},
  \citenamefont {Mihm}, \citenamefont {Müller}, \citenamefont {Müntinga},
  \citenamefont {Nepal}, \citenamefont {Oberschulte}, \citenamefont
  {Papakonstantinou}, \citenamefont {Perovsek}, \citenamefont {Peters},
  \citenamefont {Rasel}, \citenamefont {Roura}, \citenamefont {Sbroscia},
  \citenamefont {Schleich}, \citenamefont {Schubert}, \citenamefont {Seidel},
  \citenamefont {Sommer}, \citenamefont {Spindeldreier}, \citenamefont
  {Stamper-Kurn}, \citenamefont {Stuhl}, \citenamefont {Warner}, \citenamefont
  {Wendrich}, \citenamefont {Wenzlawski}, \citenamefont {Wicht}, \citenamefont
  {Windpassinger}, \citenamefont {Yu},\ and\ \citenamefont
  {Wörner}}]{frye2021bose}%
  \BibitemOpen
  \bibfield  {author} {\bibinfo {author} {\bibfnamefont {K.}~\bibnamefont
  {Frye}}, \bibinfo {author} {\bibfnamefont {S.}~\bibnamefont {Abend}},
  \bibinfo {author} {\bibfnamefont {W.}~\bibnamefont {Bartosch}}, \bibinfo
  {author} {\bibfnamefont {A.}~\bibnamefont {Bawamia}}, \bibinfo {author}
  {\bibfnamefont {D.}~\bibnamefont {Becker}}, \bibinfo {author} {\bibfnamefont
  {H.}~\bibnamefont {Blume}}, \bibinfo {author} {\bibfnamefont
  {C.}~\bibnamefont {Braxmaier}}, \bibinfo {author} {\bibfnamefont {S.-W.}\
  \bibnamefont {Chiow}}, \bibinfo {author} {\bibfnamefont {M.~A.}\ \bibnamefont
  {Efremov}}, \bibinfo {author} {\bibfnamefont {W.}~\bibnamefont {Ertmer}},
  \bibinfo {author} {\bibfnamefont {P.}~\bibnamefont {Fierlinger}}, \bibinfo
  {author} {\bibfnamefont {T.}~\bibnamefont {Franz}}, \bibinfo {author}
  {\bibfnamefont {N.}~\bibnamefont {Gaaloul}}, \bibinfo {author} {\bibfnamefont
  {J.}~\bibnamefont {Grosse}}, \bibinfo {author} {\bibfnamefont
  {C.}~\bibnamefont {Grzeschik}}, \bibinfo {author} {\bibfnamefont
  {O.}~\bibnamefont {Hellmig}}, \bibinfo {author} {\bibfnamefont {V.~A.}\
  \bibnamefont {Henderson}}, \bibinfo {author} {\bibfnamefont {W.}~\bibnamefont
  {Herr}}, \bibinfo {author} {\bibfnamefont {U.}~\bibnamefont {Israelsson}},
  \bibinfo {author} {\bibfnamefont {J.}~\bibnamefont {Kohel}}, \bibinfo
  {author} {\bibfnamefont {M.}~\bibnamefont {Krutzik}}, \bibinfo {author}
  {\bibfnamefont {C.}~\bibnamefont {Kürbis}}, \bibinfo {author} {\bibfnamefont
  {C.}~\bibnamefont {Lämmerzahl}}, \bibinfo {author} {\bibfnamefont
  {M.}~\bibnamefont {List}}, \bibinfo {author} {\bibfnamefont {D.}~\bibnamefont
  {Lüdtke}}, \bibinfo {author} {\bibfnamefont {N.}~\bibnamefont {Lundblad}},
  \bibinfo {author} {\bibfnamefont {J.~P.}\ \bibnamefont {Marburger}}, \bibinfo
  {author} {\bibfnamefont {M.}~\bibnamefont {Meister}}, \bibinfo {author}
  {\bibfnamefont {M.}~\bibnamefont {Mihm}}, \bibinfo {author} {\bibfnamefont
  {H.}~\bibnamefont {Müller}}, \bibinfo {author} {\bibfnamefont
  {H.}~\bibnamefont {Müntinga}}, \bibinfo {author} {\bibfnamefont {A.~M.}\
  \bibnamefont {Nepal}}, \bibinfo {author} {\bibfnamefont {T.}~\bibnamefont
  {Oberschulte}}, \bibinfo {author} {\bibfnamefont {A.}~\bibnamefont
  {Papakonstantinou}}, \bibinfo {author} {\bibfnamefont {J.}~\bibnamefont
  {Perovsek}}, \bibinfo {author} {\bibfnamefont {A.}~\bibnamefont {Peters},
  \bibfnamefont {Achim ;~Prat}}, \bibinfo {author} {\bibfnamefont {E.~M.}\
  \bibnamefont {Rasel}}, \bibinfo {author} {\bibfnamefont {A.}~\bibnamefont
  {Roura}}, \bibinfo {author} {\bibfnamefont {M.}~\bibnamefont {Sbroscia}},
  \bibinfo {author} {\bibfnamefont {W.~P.}\ \bibnamefont {Schleich}}, \bibinfo
  {author} {\bibfnamefont {C.}~\bibnamefont {Schubert}}, \bibinfo {author}
  {\bibfnamefont {S.~T.}\ \bibnamefont {Seidel}}, \bibinfo {author}
  {\bibfnamefont {J.}~\bibnamefont {Sommer}}, \bibinfo {author} {\bibfnamefont
  {C.}~\bibnamefont {Spindeldreier}}, \bibinfo {author} {\bibfnamefont
  {D.}~\bibnamefont {Stamper-Kurn}}, \bibinfo {author} {\bibfnamefont {B.~K.}\
  \bibnamefont {Stuhl}}, \bibinfo {author} {\bibfnamefont {M.}~\bibnamefont
  {Warner}}, \bibinfo {author} {\bibfnamefont {T.}~\bibnamefont {Wendrich}},
  \bibinfo {author} {\bibfnamefont {A.}~\bibnamefont {Wenzlawski}}, \bibinfo
  {author} {\bibfnamefont {A.}~\bibnamefont {Wicht}}, \bibinfo {author}
  {\bibfnamefont {P.}~\bibnamefont {Windpassinger}}, \bibinfo {author}
  {\bibfnamefont {N.}~\bibnamefont {Yu}}, \ and\ \bibinfo {author}
  {\bibfnamefont {L.}~\bibnamefont {Wörner}},\ }\href@noop {} {\bibfield
  {journal} {\bibinfo  {journal} {EPJ Quantum Technology}\ }\textbf {\bibinfo
  {volume} {8}},\ \bibinfo {pages} {1} (\bibinfo {year} {2021})}\BibitemShut
  {NoStop}%
\bibitem [{\citenamefont {Bertoldi}\ \emph {et~al.}(2021)\citenamefont
  {Bertoldi}, \citenamefont {Bongs}, \citenamefont {Bouyer}, \citenamefont
  {Buchmueller}, \citenamefont {Canuel}, \citenamefont {Caramete},
  \citenamefont {Chiofalo}, \citenamefont {Coleman}, \citenamefont {De~Roeck},
  \citenamefont {Ellis}, \citenamefont {Graham}, \citenamefont {Haehnelt},
  \citenamefont {Hees}, \citenamefont {Hogan}, \citenamefont {von Klitzing},
  \citenamefont {Krutzik}, \citenamefont {Lewicki}, \citenamefont {McCabe},
  \citenamefont {Peters}, \citenamefont {Rasel}, \citenamefont {Roura},
  \citenamefont {Sabulsky}, \citenamefont {Schiller}, \citenamefont {Schubert},
  \citenamefont {Signorini}, \citenamefont {Sorrentino}, \citenamefont {Singh},
  \citenamefont {Tino}, \citenamefont {Vaskonen},\ and\ \citenamefont
  {Zhan}}]{abou2020aedge}%
  \BibitemOpen
  \bibfield  {author} {\bibinfo {author} {\bibfnamefont {A.}~\bibnamefont
  {Bertoldi}}, \bibinfo {author} {\bibfnamefont {K.}~\bibnamefont {Bongs}},
  \bibinfo {author} {\bibfnamefont {P.}~\bibnamefont {Bouyer}}, \bibinfo
  {author} {\bibfnamefont {O.}~\bibnamefont {Buchmueller}}, \bibinfo {author}
  {\bibfnamefont {B.}~\bibnamefont {Canuel}}, \bibinfo {author} {\bibfnamefont
  {L.-I.}\ \bibnamefont {Caramete}}, \bibinfo {author} {\bibfnamefont {M.~L.}\
  \bibnamefont {Chiofalo}}, \bibinfo {author} {\bibfnamefont {J.}~\bibnamefont
  {Coleman}}, \bibinfo {author} {\bibfnamefont {A.}~\bibnamefont {De~Roeck}},
  \bibinfo {author} {\bibfnamefont {J.}~\bibnamefont {Ellis}}, \bibinfo
  {author} {\bibfnamefont {P.~W.}\ \bibnamefont {Graham}}, \bibinfo {author}
  {\bibfnamefont {M.~G.}\ \bibnamefont {Haehnelt}}, \bibinfo {author}
  {\bibfnamefont {A.}~\bibnamefont {Hees}}, \bibinfo {author} {\bibfnamefont
  {J.}~\bibnamefont {Hogan}}, \bibinfo {author} {\bibfnamefont
  {W.}~\bibnamefont {von Klitzing}}, \bibinfo {author} {\bibfnamefont
  {M.}~\bibnamefont {Krutzik}}, \bibinfo {author} {\bibfnamefont
  {M.}~\bibnamefont {Lewicki}}, \bibinfo {author} {\bibfnamefont
  {C.}~\bibnamefont {McCabe}}, \bibinfo {author} {\bibfnamefont
  {A.}~\bibnamefont {Peters}}, \bibinfo {author} {\bibfnamefont
  {E.}~\bibnamefont {Rasel}}, \bibinfo {author} {\bibfnamefont
  {A.}~\bibnamefont {Roura}}, \bibinfo {author} {\bibfnamefont
  {D.}~\bibnamefont {Sabulsky}}, \bibinfo {author} {\bibfnamefont
  {S.}~\bibnamefont {Schiller}}, \bibinfo {author} {\bibfnamefont
  {C.}~\bibnamefont {Schubert}}, \bibinfo {author} {\bibfnamefont
  {C.}~\bibnamefont {Signorini}}, \bibinfo {author} {\bibfnamefont
  {F.}~\bibnamefont {Sorrentino}}, \bibinfo {author} {\bibfnamefont
  {Y.}~\bibnamefont {Singh}}, \bibinfo {author} {\bibfnamefont {G.~M.}\
  \bibnamefont {Tino}}, \bibinfo {author} {\bibfnamefont {V.}~\bibnamefont
  {Vaskonen}}, \ and\ \bibinfo {author} {\bibfnamefont {M.-S.}\ \bibnamefont
  {Zhan}},\ }\href@noop {} {\bibfield  {journal} {\bibinfo  {journal}
  {Experimental Astronomy}\ }\textbf {\bibinfo {volume} {51}},\ \bibinfo
  {pages} {1417} (\bibinfo {year} {2021})}\BibitemShut {NoStop}%
\bibitem [{\citenamefont {Elliott}\ \emph {et~al.}(2018)\citenamefont
  {Elliott}, \citenamefont {Krutzik}, \citenamefont {Williams}, \citenamefont
  {Thompson},\ and\ \citenamefont {Aveline}}]{elliott2018nasa}%
  \BibitemOpen
  \bibfield  {author} {\bibinfo {author} {\bibfnamefont {E.~R.}\ \bibnamefont
  {Elliott}}, \bibinfo {author} {\bibfnamefont {M.~C.}\ \bibnamefont
  {Krutzik}}, \bibinfo {author} {\bibfnamefont {J.~R.}\ \bibnamefont
  {Williams}}, \bibinfo {author} {\bibfnamefont {R.~J.}\ \bibnamefont
  {Thompson}}, \ and\ \bibinfo {author} {\bibfnamefont {D.~C.}\ \bibnamefont
  {Aveline}},\ }\href@noop {} {\bibfield  {journal} {\bibinfo  {journal} {NPJ
  Microgravity}\ }\textbf {\bibinfo {volume} {4}},\ \bibinfo {pages} {1}
  (\bibinfo {year} {2018})}\BibitemShut {NoStop}%
\bibitem [{\citenamefont {Cooper}\ \emph
  {et~al.}(2021{\natexlab{b}})\citenamefont {Cooper}, \citenamefont {Madkhaly},
  \citenamefont {Johnson}, \citenamefont {Baldolini},\ and\ \citenamefont
  {Hackerm{\"u}ller}}]{cooper2021dual}%
  \BibitemOpen
  \bibfield  {author} {\bibinfo {author} {\bibfnamefont {N.}~\bibnamefont
  {Cooper}}, \bibinfo {author} {\bibfnamefont {S.}~\bibnamefont {Madkhaly}},
  \bibinfo {author} {\bibfnamefont {D.}~\bibnamefont {Johnson}}, \bibinfo
  {author} {\bibfnamefont {D.}~\bibnamefont {Baldolini}}, \ and\ \bibinfo
  {author} {\bibfnamefont {L.}~\bibnamefont {Hackerm{\"u}ller}},\ }\href@noop
  {} {\bibfield  {journal} {\bibinfo  {journal} {arXiv preprint
  arXiv:2106.11014}\ } (\bibinfo {year} {2021}{\natexlab{b}})}\BibitemShut
  {NoStop}%
\bibitem [{\citenamefont {Ultimaker}(2020)}]{ultimaker}%
  \BibitemOpen
  \bibfield  {author} {\bibinfo {author} {\bibnamefont {Ultimaker}},\
  }\href@noop {} {\enquote {\bibinfo {title} {The widest material choice on the
  market},}\ } (\bibinfo {year} {2011-2020})\BibitemShut {NoStop}%
\bibitem [{\citenamefont {Simplify3D}(2021)}]{simplify3d}%
  \BibitemOpen
  \bibfield  {author} {\bibinfo {author} {\bibnamefont {Simplify3D}},\
  }\href@noop {} {\enquote {\bibinfo {title} {Filament properties table},}\ }
  (\bibinfo {year} {2021})\BibitemShut {NoStop}%
\bibitem [{\citenamefont {Friedrich}\ and\ \citenamefont
  {Walter}(2020)}]{friedrich2020structure}%
  \BibitemOpen
  \bibfield  {author} {\bibinfo {author} {\bibfnamefont {K.}~\bibnamefont
  {Friedrich}}\ and\ \bibinfo {author} {\bibfnamefont {R.}~\bibnamefont
  {Walter}},\ }\href@noop {} {\emph {\bibinfo {title} {Structure and Properties
  of Additive Manufactured Polymer Components}}}\ (\bibinfo  {publisher}
  {Woodhead Publishing},\ \bibinfo {year} {2020})\BibitemShut {NoStop}%
\bibitem [{\citenamefont {Preston}(1996)}]{sas}%
  \BibitemOpen
  \bibfield  {author} {\bibinfo {author} {\bibfnamefont {D.~W.}\ \bibnamefont
  {Preston}},\ }\href@noop {} {\bibfield  {journal} {\bibinfo  {journal}
  {American Journal of Physics}\ }\textbf {\bibinfo {volume} {64}},\ \bibinfo
  {pages} {1432} (\bibinfo {year} {1996})}\BibitemShut {NoStop}%
\bibitem [{\citenamefont {Goh}\ \emph {et~al.}(2020)\citenamefont {Goh},
  \citenamefont {Yap}, \citenamefont {Tan}, \citenamefont {Sing}, \citenamefont
  {Goh},\ and\ \citenamefont {Yeong}}]{goh2020process}%
  \BibitemOpen
  \bibfield  {author} {\bibinfo {author} {\bibfnamefont {G.~D.}\ \bibnamefont
  {Goh}}, \bibinfo {author} {\bibfnamefont {Y.~L.}\ \bibnamefont {Yap}},
  \bibinfo {author} {\bibfnamefont {H.}~\bibnamefont {Tan}}, \bibinfo {author}
  {\bibfnamefont {S.~L.}\ \bibnamefont {Sing}}, \bibinfo {author}
  {\bibfnamefont {G.~L.}\ \bibnamefont {Goh}}, \ and\ \bibinfo {author}
  {\bibfnamefont {W.~Y.}\ \bibnamefont {Yeong}},\ }\href@noop {} {\bibfield
  {journal} {\bibinfo  {journal} {Crit. Rev. Solid State Mater. Sci.}\ }\textbf
  {\bibinfo {volume} {45}},\ \bibinfo {pages} {113} (\bibinfo {year}
  {2020})}\BibitemShut {NoStop}%
\bibitem [{\citenamefont {Duncker}\ \emph {et~al.}(2014)\citenamefont
  {Duncker}, \citenamefont {Hellmig}, \citenamefont {Wenzlawski}, \citenamefont
  {Grote}, \citenamefont {Rafipoor}, \citenamefont {Rafipoor}, \citenamefont
  {Sengstock},\ and\ \citenamefont {Windpassinger}}]{duncker2014ultrastable}%
  \BibitemOpen
  \bibfield  {author} {\bibinfo {author} {\bibfnamefont {H.}~\bibnamefont
  {Duncker}}, \bibinfo {author} {\bibfnamefont {O.}~\bibnamefont {Hellmig}},
  \bibinfo {author} {\bibfnamefont {A.}~\bibnamefont {Wenzlawski}}, \bibinfo
  {author} {\bibfnamefont {A.}~\bibnamefont {Grote}}, \bibinfo {author}
  {\bibfnamefont {A.~J.}\ \bibnamefont {Rafipoor}}, \bibinfo {author}
  {\bibfnamefont {M.}~\bibnamefont {Rafipoor}}, \bibinfo {author}
  {\bibfnamefont {K.}~\bibnamefont {Sengstock}}, \ and\ \bibinfo {author}
  {\bibfnamefont {P.}~\bibnamefont {Windpassinger}},\ }\href@noop {} {\bibfield
   {journal} {\bibinfo  {journal} {Appl. Opt}\ }\textbf {\bibinfo {volume}
  {53}},\ \bibinfo {pages} {4468} (\bibinfo {year} {2014})}\BibitemShut
  {NoStop}%
\bibitem [{\citenamefont {Ressel}\ \emph {et~al.}(2010)\citenamefont {Ressel},
  \citenamefont {Gohlke}, \citenamefont {Rauen}, \citenamefont {Schuldt},
  \citenamefont {Kronast}, \citenamefont {Mescheder}, \citenamefont {Johann},
  \citenamefont {Weise},\ and\ \citenamefont
  {Braxmaier}}]{ressel2010ultrastable}%
  \BibitemOpen
  \bibfield  {author} {\bibinfo {author} {\bibfnamefont {S.}~\bibnamefont
  {Ressel}}, \bibinfo {author} {\bibfnamefont {M.}~\bibnamefont {Gohlke}},
  \bibinfo {author} {\bibfnamefont {D.}~\bibnamefont {Rauen}}, \bibinfo
  {author} {\bibfnamefont {T.}~\bibnamefont {Schuldt}}, \bibinfo {author}
  {\bibfnamefont {W.}~\bibnamefont {Kronast}}, \bibinfo {author} {\bibfnamefont
  {U.}~\bibnamefont {Mescheder}}, \bibinfo {author} {\bibfnamefont
  {U.}~\bibnamefont {Johann}}, \bibinfo {author} {\bibfnamefont
  {D.}~\bibnamefont {Weise}}, \ and\ \bibinfo {author} {\bibfnamefont
  {C.}~\bibnamefont {Braxmaier}},\ }\href@noop {} {\bibfield  {journal}
  {\bibinfo  {journal} {Appl. Opt}\ }\textbf {\bibinfo {volume} {49}},\
  \bibinfo {pages} {4296} (\bibinfo {year} {2010})}\BibitemShut {NoStop}%
\bibitem [{\citenamefont {Chonhenchob}\ \emph {et~al.}(2012)\citenamefont
  {Chonhenchob}, \citenamefont {Singh}, \citenamefont {Singh}, \citenamefont
  {Stallings},\ and\ \citenamefont {Grewal}}]{chonhenchob2012measurement}%
  \BibitemOpen
  \bibfield  {author} {\bibinfo {author} {\bibfnamefont {V.}~\bibnamefont
  {Chonhenchob}}, \bibinfo {author} {\bibfnamefont {S.~P.}\ \bibnamefont
  {Singh}}, \bibinfo {author} {\bibfnamefont {J.~J.}\ \bibnamefont {Singh}},
  \bibinfo {author} {\bibfnamefont {J.}~\bibnamefont {Stallings}}, \ and\
  \bibinfo {author} {\bibfnamefont {G.}~\bibnamefont {Grewal}},\ }\href@noop {}
  {\bibfield  {journal} {\bibinfo  {journal} {Packaging Technology and
  Science}\ }\textbf {\bibinfo {volume} {25}},\ \bibinfo {pages} {31} (\bibinfo
  {year} {2012})}\BibitemShut {NoStop}%
\bibitem [{\citenamefont {Schkolnik}\ \emph {et~al.}(2016)\citenamefont
  {Schkolnik}, \citenamefont {Hellmig}, \citenamefont {Wenzlawski},
  \citenamefont {Grosse}, \citenamefont {Kohfeldt}, \citenamefont
  {D{\"o}ringshoff}, \citenamefont {Wicht}, \citenamefont {Windpassinger},
  \citenamefont {Sengstock}, \citenamefont {Braxmaier}, \citenamefont
  {Krutzik},\ and\ \citenamefont {Peters}}]{schkolnik2016compact}%
  \BibitemOpen
  \bibfield  {author} {\bibinfo {author} {\bibfnamefont {V.}~\bibnamefont
  {Schkolnik}}, \bibinfo {author} {\bibfnamefont {O.}~\bibnamefont {Hellmig}},
  \bibinfo {author} {\bibfnamefont {A.}~\bibnamefont {Wenzlawski}}, \bibinfo
  {author} {\bibfnamefont {J.}~\bibnamefont {Grosse}}, \bibinfo {author}
  {\bibfnamefont {A.}~\bibnamefont {Kohfeldt}}, \bibinfo {author}
  {\bibfnamefont {K.}~\bibnamefont {D{\"o}ringshoff}}, \bibinfo {author}
  {\bibfnamefont {A.}~\bibnamefont {Wicht}}, \bibinfo {author} {\bibfnamefont
  {P.}~\bibnamefont {Windpassinger}}, \bibinfo {author} {\bibfnamefont
  {K.}~\bibnamefont {Sengstock}}, \bibinfo {author} {\bibfnamefont
  {C.}~\bibnamefont {Braxmaier}}, \bibinfo {author} {\bibfnamefont
  {M.}~\bibnamefont {Krutzik}}, \ and\ \bibinfo {author} {\bibfnamefont
  {A.}~\bibnamefont {Peters}},\ }\href@noop {} {\bibfield  {journal} {\bibinfo
  {journal} {Applied Physics B}\ }\textbf {\bibinfo {volume} {122}},\ \bibinfo
  {pages} {1} (\bibinfo {year} {2016})}\BibitemShut {NoStop}%
\end{thebibliography}%



\end{document}